\documentclass[11pt]{article}

\usepackage[a4paper,margin=25mm]{geometry}
\usepackage{amsmath,amssymb}
\usepackage{graphicx}
\usepackage{float}
\usepackage{booktabs,array}
\usepackage{setspace}
\usepackage[hidelinks]{hyperref}

\newcolumntype{L}[1]{>{\raggedright\arraybackslash}p{#1}}

\begin{document}

\begin{center}
{\Large\bfseries Residual diagnostics for detecting the breakdown of the single power-law approximation in photoelectron yield spectroscopy: application to doped silicon\par}
\vspace{5mm}
Shinjiro Yagyu$^{a,*}$, Takahiro Nagata$^{a}$, and Yoshiyuki Nakajima$^{b}$\par
\vspace{3mm}
{\small
$^{a}$Functional Materials Field, Research Center for Electronic and Optical Materials, National Institute for Materials Science, 1-1 Namiki, Tsukuba, Ibaraki 305-0044, Japan\par
$^{b}$RIKEN KEIKI Co., Ltd., Itabashi-ku, Tokyo 174-8744, Japan\par
$^{*}$Corresponding author: YAGYU.Shinjiro@nims.go.jp\par}
\end{center}

\begin{abstract}
Our automated analysis of photoelectron yield spectroscopy (PYS) estimates the threshold $I_{th}$ and the exponent $n$ of a single power law for every spectrum. When this model fits poorly, however, the analysis cannot tell whether the data or the model is at fault. Here we present residual diagnostics that separate the two causes. The exponent is scanned continuously on a $1/n$-power fit with an Akaike information criterion (AIC) based on the mean absolute error (MAE), and the spread of the resulting Akaike weights ($P \times \Delta n$) quantifies the uncertainty of the estimate. The normalized MAE (NMAE) and the RMSE-to-MAE ratio (RMR) assess the data quality, the Durbin--Watson (DW) statistic assesses the model adequacy, and the three checks are reported in parallel. With criteria set from an Au reference, the diagnostics were applied to 60 continuous measurements in air of four HF-treated Si samples. For heavily doped p-type Si, DW fell below its criterion, whereas the means of NMAE and RMR remained within their criteria. Prompted by this flag, we fitted the spectrum with a two-component model with fixed exponents ($n = 1$ and 2.5); the AIC decreased by 56, DW increased from 0.86 to 1.99, and the two components set in at 4.70 and 5.28~eV. For heavily doped n-type Si, the change during air exposure was consistent with a single component. The diagnostics thus identify the spectra that require a multi-component analysis without assuming an emission mechanism.
\end{abstract}

\noindent\textbf{Keywords:} photoelectron yield spectroscopy; residual diagnostics; Durbin--Watson statistic; Akaike information criterion; silicon; automated analysis

\section{Introduction}

Photoelectron yield spectroscopy (PYS) determines the ionization energy of semiconductors and the work function of metals from the threshold energy $I_{th}$ of the photoemission yield $Y$, which rises with the photon energy $h\nu$ as a power law, $Y \propto (h\nu - I_{th})^n$. Kane's theory for semiconductors showed that the exponent $n$ reflects the excitation process, the scattering of photoelectrons before escape, and the origin of the emission~\cite{Kane1962}: $n = 1$ for an unscattered direct transition in the volume, $n = 2$ when the directly excited electron is elastically scattered, $n = 5/2$ for an indirect (phonon- or imperfection-assisted) transition, and $n = 3/2$ for a direct transition from surface band states whose threshold coincides with the Fermi level (see Table~S6 of the Supplementary Material for the full list). Gobeli and Allen~\cite{Gobeli1962} analyzed the spectrum of a cleaved Si surface in ultrahigh vacuum as the sum of a direct and an indirect transition, $Y = A_1(h\nu - I_{th,1})^{n_1} + A_2(h\nu - I_{th,2})^{n_2}$, with $n_1 = 1$ for the direct transition and Kane's $n_2 = 5/2$ for the indirect transition, whose threshold lies about 0.3~eV below that of the direct one. They also noted that the indirect component agreed with an empirical cube law within their experimental accuracy~\cite{Gobeli1962}, and $n = 3$ is also used in PYS analyses of Si~\cite{Miyazaki1999}. This two-component description remains a reference for the PYS analysis of Si-based materials.

We have developed automated methods for analyzing PYS spectra. In our first method~\cite{Yagyu2025}, piecewise regression was used to extract the region that follows a power law, and for an exponent specified by the analyst (e.g., $n = 2$ or 3), the yield raised to the power $1/n$ was fitted with an extended rectified linear unit (exReLU) function by minimizing the mean absolute error (MAE). In the second~\cite{Yagyu2026}, candidate exponents were estimated from the data by a shift-corrected log-log linearization (3LS), and an Akaike information criterion (AIC) based on the MAE in the original space was used to select among them. Both methods return $n$ and $I_{th}$ for any spectrum. When the spectrum is a sum of components with different thresholds or exponents, however, a single power law is inadequate, and the estimates are only effective values for the sum. The estimates alone do not reveal such cases.

Automated fitting routines commonly accept or reject a result on the basis of a single goodness-of-fit value such as $R^2$. Because $R^2$ reflects only the magnitude of the residuals, a low value does not reveal its cause: noisy data, isolated spikes, or a mismatch between the model and the spectrum. In autonomous experimental systems, where the analysis of one measurement guides the choice of the next~\cite{Miyawaki2026}, the appropriate action depends on this cause: repeating the measurement, discarding a spike, or moving to a multi-component analysis.

In this work, we add residual diagnostics of the single power-law fit to our automated analysis. First, extending our previous methods~\cite{Yagyu2025,Yagyu2026}, we introduce the $1/n$-Scan, which scans the exponent continuously on the $1/n$-power fit; the dependence of the AIC on $n$ obtained in the scan quantifies the uncertainty of the estimate through the Akaike weights. Second, we evaluate the data quality and the model adequacy with residual metrics and report three checks (estimation, data quality, and model adequacy) in parallel, together with the underlying values. The model-adequacy check uses the Durbin--Watson (DW) statistic, which measures the serial correlation of the residuals; for continuous measurements, a moving median of DW over recent measurements suppresses false flags caused by the scatter of individual spectra. We set the criteria from a polycrystalline Au reference and apply the diagnostics to continuous measurements in air of four doped Si samples, in which band bending and the overlap of emission components vary with doping. Spectra flagged by the model-adequacy check are then examined with a two-component fit.

\section{Methods}

\subsection{Estimation of the effective exponent by the \texorpdfstring{$1/n$}{1/n}-Scan}

As in our previous reports~\cite{Yagyu2025,Yagyu2026}, the yield is modeled as
\begin{equation}
Y = \begin{cases} Y_{bg} & (h\nu < I_{th}) \\ A(h\nu - I_{th})^n + Y_{bg} & (h\nu \ge I_{th}) \end{cases}
\label{eq:model}
\end{equation}
In our first method~\cite{Yagyu2025}, the yield raised to the power $1/n$ for a specified $n$ was fitted by minimizing the MAE with the exReLU function
\begin{equation}
f(h\nu) = \begin{cases} B' & (h\nu < I_{th}) \\ A'(h\nu - I_{th}) + B' & (h\nu \ge I_{th}) \end{cases}
\label{eq:exrelu}
\end{equation}
In the second~\cite{Yagyu2026}, the candidates for $n$ were estimated by 3LS, which uses the logarithm of the yield and therefore becomes unstable when the noise is large~\cite{Yagyu2026}. Here, we avoid the logarithmic transformation and scan $n$ continuously on the $1/n$-power fit ($1/n$-Scan).

For each $n$, the fitted values in the original space are $\hat{Y} = f^n$, that is, $\hat{Y} = [A'(h\nu - I_{th}) + B']^n$ above the threshold and $B'^n$ below it. When $B'$ is small compared with $A'(h\nu - I_{th})$, this approximates Eq.~(\ref{eq:model}) with $A = A'^n$ and $Y_{bg} = B'^n$. The background $B'$ is a free parameter with a lower bound of zero. From the residuals $e_i = Y_i - \hat{Y}_i$ in the original space, the AIC is computed under a Laplace error distribution~\cite{Yagyu2026},
\begin{equation}
\mathrm{AIC} = 2N\ln(2\cdot\mathrm{MAE}) + 2N + 2k,
\label{eq:aic}
\end{equation}
where $N$ is the number of data points and $k = 3$ is the number of structural parameters ($I_{th}$, $A'$, and $B'$). The error scale, which is common to all models, is not counted in $k$. Our previous work~\cite{Yagyu2026} counted it; this shifts all AIC values by a constant and leaves the differences between models unchanged. The exponent $n$ is scanned over $[0.5, 4.0]$ with a step width $\Delta n$ of 0.01--0.05, and the $n$ that minimizes the AIC is taken as the estimate. With $\Delta n = 0.01$, the scan takes about 4~s per spectrum (simulated spectrum with $N = 45$, single CPU thread). Initial values of the threshold and the background are obtained by piecewise linear regression, and the fit covers the full measured range.

We refer to the estimated $n$ as the effective exponent. It equals a theoretical exponent only when a single process contributes to the spectrum; otherwise, it describes the best single power law for the sum of the components.

\subsection{Uncertainty of the estimate}

With $\Delta\mathrm{AIC}_i = \mathrm{AIC}_i - \mathrm{AIC}_{\min}$, the AIC values of the scanned exponents are converted into Akaike weights,
\begin{equation}
w_i = \frac{\exp(-\Delta\mathrm{AIC}_i/2)}{\sum_j \exp(-\Delta\mathrm{AIC}_j/2)},
\end{equation}
which we treat as a probability distribution over $n$. The perplexity $P = 2^H$, where $H = -\sum_i w_i \log_2 w_i$ is the Shannon entropy, gives the effective number of exponents supported by the data, and the product $P \times \Delta n$ serves as the effective width of the estimate. Because $P$ increases approximately in inverse proportion to the step width $\Delta n$, the product is independent of the step width. The exponents of Kane's theory (1, 3/2, 2, and 5/2) are spaced by 0.5 (Table~S6). An exponent can be assigned to a single process when this width is below half the spacing (0.25); allowing a margin for measurement error, we set the criterion to $P \times \Delta n \le 0.3$ and use the full spacing, 0.5, as the upper warning level. For material systems with a different spacing of exponents, the criterion is derived from that spacing.

\subsection{Residual metrics}

Four metrics are computed from the residuals of the selected fit, with $\mathrm{MAE} = \frac{1}{N}\sum|e_i|$ and $\mathrm{RMSE} = \sqrt{\frac{1}{N}\sum e_i^2}$.

\textbf{NMAE (data quality).} The MAE normalized by the mean yield $\bar{Y}$,
\begin{equation}
\mathrm{NMAE} = \frac{\mathrm{MAE}}{\bar{Y}} \times 100\ [\%],
\end{equation}
measures the overall noise relative to the signal. A high NMAE indicates an insufficient signal or a noisy measurement.

\textbf{RMR (isolated anomalies).} The ratio of RMSE to MAE, normalized by its value for Gaussian noise,
\begin{equation}
\mathrm{RMR} = \frac{\mathrm{RMSE}}{\mathrm{MAE}} \bigg/ \sqrt{\frac{\pi}{2}},
\end{equation}
is 1.00 for uniform Gaussian noise and $2/\sqrt{\pi} \approx 1.13$ for Laplace noise, and it increases when isolated spikes are present: because the fit minimizes the MAE, a spike remains in the residuals and raises the RMSE more than the MAE. Shot noise, which grows with the yield, also raises RMR above 1, because the large residuals concentrate on the high-yield side (Supplementary Material S6). Our previous work~\cite{Yagyu2026} used the unnormalized ratio; an RMR of 2.0 here corresponds to an unnormalized value of 2.5, the lower bound of the warning band defined there.

\textbf{DW (model adequacy).} The Durbin--Watson statistic
\begin{equation}
\mathrm{DW} = \frac{\sum_{i=2}^{N}(e_i - e_{i-1})^2}{\sum_{i=1}^{N} e_i^2}
\end{equation}
measures the first-order autocorrelation $r$ of the residuals through the differences between adjacent residuals ($\mathrm{DW} \approx 2(1 - r)$). It is close to 2 for uncorrelated residuals and decreases when the residuals undulate along the energy axis. When components with different thresholds overlap, a single power law deviates smoothly from the spectrum, and the residuals show positive first-order correlation even when NMAE and RMR are within their criteria. As the criterion, we use the lower bound of the 5\% one-sided test for positive autocorrelation, $d_L \approx 1.46$ ($N \approx 50$)~\cite{Durbin1951,Savin1977}. DW is computed over all $N$ points of the measured range. Because the fitted threshold changes with $n$, restricting the calculation to the points above the threshold would make the number of points depend on $n$ and on the sample; over the full range, $N$ and $d_L$ are the same for all $n$ and all samples, and any mismatch near the onset is included. The assumptions behind $d_L$ and the distribution of DW under shot noise are given in Supplementary Material S6.

\textbf{$\Delta R^2$ (extent of the deviation).} The coefficients of determination based on the squared and absolute errors,
\begin{equation}
R^2_{\mathrm{RMSE}} = 1 - \frac{\sum e_i^2}{\sum (Y_i - \bar{Y})^2}, \qquad
R^2_{\mathrm{MAE}} = 1 - \frac{\sum |e_i|}{\sum |Y_i - \tilde{Y}|}, \qquad
\Delta R^2 = |R^2_{\mathrm{RMSE}} - R^2_{\mathrm{MAE}}|,
\end{equation}
where $\tilde{Y}$ is the median of the yield, differ in their sensitivity to large residuals. $\Delta R^2$ increases when large residuals concentrate in the high-yield region, as happens when the model deviates on the high-energy side. Because it responds to the magnitude of the residuals but not to their order along the energy axis, $\Delta R^2$ is used only as an auxiliary indicator of model adequacy.

The residuals are not standardized ($e_i/\sigma_i$), because estimating $\sigma_i$ point by point in spectra of 50 or fewer points adds scatter to the metrics. Even when $\sigma_i$ varies between points, as for shot noise, the mean of DW remains close to 2 as long as the residuals are independent (Supplementary Material S6).

\subsection{Combination of checks and series rule}

\begin{figure}[p]
\centering
\includegraphics[width=\linewidth,height=0.75\textheight,keepaspectratio]{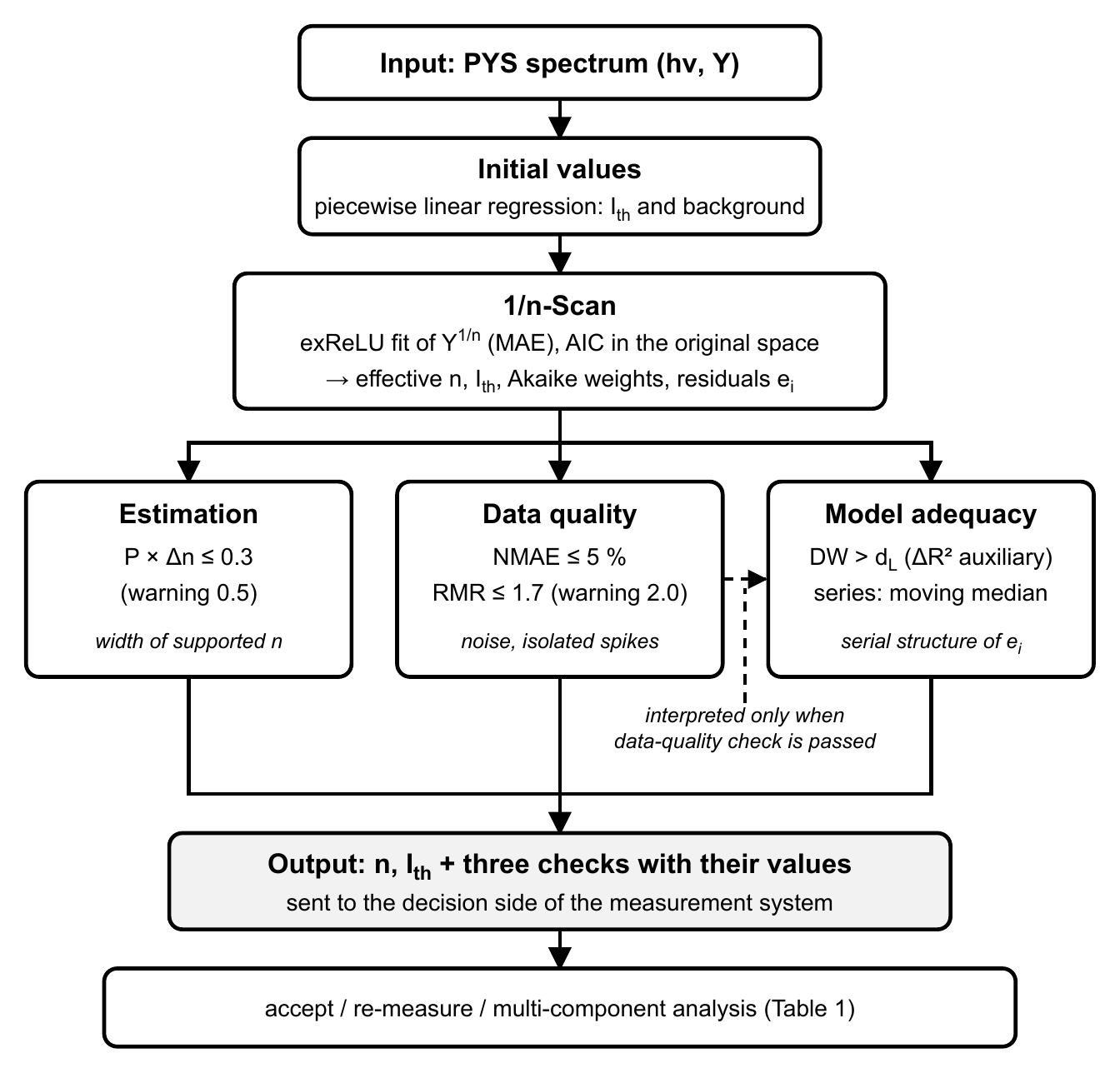}
\caption{Structure of the diagnosis. Initial values are obtained by piecewise linear regression, and the effective exponent and the threshold are estimated by the $1/n$-Scan ($1/n$-power fit with exReLU and MAE~\cite{Yagyu2025}, AIC in the original space~\cite{Yagyu2026}, scanned over $n$). Three checks, estimation ($P \times \Delta n$), data quality (NMAE and RMR), and model adequacy (DW, with $\Delta R^2$ as an auxiliary indicator and the moving median of DW for continuous measurements), are evaluated in parallel and reported together with their values. The model-adequacy check is interpreted only when the data-quality check is passed (dashed line). The interpretation of the combinations is given in Table~\ref{tab:decision}.}
\label{fig:flow}
\end{figure}

The three checks, estimation ($P \times \Delta n$), data quality (NMAE and RMR), and model adequacy (DW, with $\Delta R^2$ as an auxiliary indicator), are evaluated for every spectrum and reported together with their values (Fig.~\ref{fig:flow}). Table~\ref{tab:decision} gives the interpretation of their combinations. The checks have one dependency: because DW also decreases for noisy data, the model-adequacy check is interpreted only when the data-quality check is passed. The estimation check is read together with the other two as information on the range of supported exponents, because an overlap of components also broadens the AIC minimum.

For continuous measurements, to suppress the scatter of DW between individual spectra, the model-adequacy check is flagged when the moving median of DW over the last ten measurements falls below $d_L$. Under shot noise, 14\% of individual spectra that follow a single power law give DW below $d_L$; with a window of ten measurements, the probability that the rule flags such spectra is below $10^{-3}$ (Supplementary Material S6).

\subsection{Example of the diagnostic output}

Figure~\ref{fig:example} shows the diagnostic output for the 43rd spectrum of the HF-treated NL sample. The left column shows the result of the $1/n$-Scan: (a) the AIC and (b) the Akaike weights, which determine the estimation check; (c) the threshold for each $n$; and (d)--(g) the residual metrics computed for each $n$. The right column shows the best fit and its residuals at the selected $n$ in the $1/n$ space (h, i) and in the original space (j, k).

\section{Experimental}

Four Si(100) wafers were used: p-type high resistivity (PH), p-type low resistivity (PL), n-type high resistivity (NH), and n-type low resistivity (NL). The resistivity of PL and NL ($<0.02~\Omega$\,cm) corresponds to impurity concentrations on the order of $10^{18}$--$10^{19}$~cm$^{-3}$. Each sample was cleaned ultrasonically in acetone and ethanol, immersed in HF solution for 2~min, and rinsed with pure water. The contact potential difference (CPD) and PYS measurements were started within 3~h of the HF treatment.

PYS spectra were measured in air with a RIKEN KEIKI AC-5 from 3.8 to 6.2~eV in 0.05~eV steps ($N = 49$, about 8~min per spectrum). Each Si sample was measured continuously (PL, PH, and NH: 99 times; NL: 61 times), and the first 60 measurements were analyzed. As a single-component reference, a rolled Au plate (99.999\%) was measured continuously 10 times under the same conditions. The CPD was measured in air before and after each PYS series with a RIKEN KEIKI FAC-2 Kelvin probe calibrated against the PYS threshold of a rolled Au plate. X-ray photoelectron spectra (XPS) of separately HF-treated samples were recorded with a ULVAC-PHI Quantera (monochromated Al K$\alpha$). Details are given in Supplementary Material S1 and S2.

\section{Results}

\subsection{Criteria from the Au reference}

For the 10 measurements of Au, the mean values were $P \times \Delta n = 0.27$, NMAE $= 4.5\%$, RMR $= 1.62$, DW $= 2.15$ (median 2.06), and $\Delta R^2 = 0.042$ (0.030--0.058) (Supplementary Material S3, Figs.~S3 and S4). The residuals were close to white noise, and a single power law described the spectra within the noise. The exponent was 1.65--2.02 (mean 1.84), slightly below the Fowler value of $n = 2$ for metals~\cite{Fowler1931}. We set the criteria empirically 5--20\% above these means: NMAE $\le 5\%$, RMR $\le 1.7$ (upper warning level 2.0), $\Delta R^2 \le 0.05$, and $P \times \Delta n \le 0.3$. The DW criterion is $d_L \approx 1.46$. Because the criteria were derived from the means, some individual Au spectra exceeded them: NMAE in 2 and RMR in 2 of the 10 measurements. DW was above $d_L$ in all 10. The Si samples were evaluated with these criteria (see Supplementary Material S6 for the derivation).

\subsection{Diagnosis of the doped Si series}

\begin{figure}[p]
\centering
\includegraphics[width=0.98\linewidth,height=0.80\textheight,keepaspectratio]{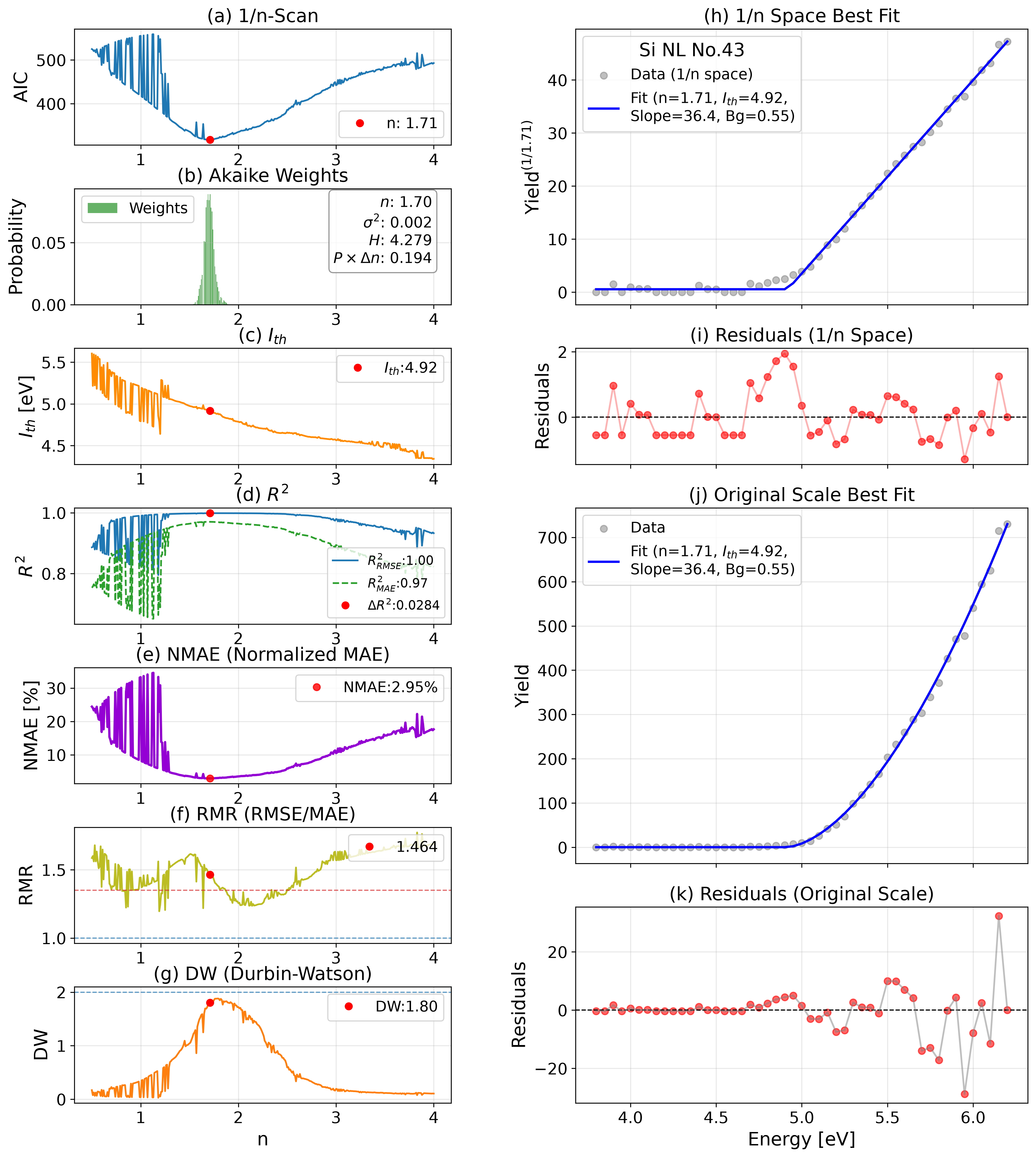}
\caption{Diagnostic output for the 43rd spectrum of the HF-treated NL sample. (a) AIC against $n$; the legend gives the $n$ with the minimum AIC. (b) Akaike weights; the box gives the mean of $n$ weighted by the Akaike weights, its variance $\sigma^2$, the entropy $H$, and $P \times \Delta n$. (c) Threshold $I_{th}$ for each $n$. (d) $R^2$ based on RMSE and MAE and their difference $\Delta R^2$. (e) NMAE (\%). (f) RMR. (g) DW. (h) Best fit in the $1/n$ space and (i) its residuals. (j) Best fit in the original space and (k) its residuals. All three checks are passed ($P \times \Delta n = 0.19$, NMAE $= 2.95\%$, RMR $= 1.46$, DW $= 1.80$). The Akaike weights have a single sharp peak near $n = 1.7$, and the residuals show no undulation. The larger residual amplitude on the high-yield side reflects shot noise. Bg in (h) and (j) is $B'$ in the $1/n$ space. The jumps of the curves in (a)--(g) occur where the fitted threshold changes between local minima; their AIC values lie above the minimum, and they do not affect the selected $n$.}
\label{fig:example}
\end{figure}

\begin{figure}[p]
\centering
\includegraphics[width=\linewidth,height=0.80\textheight,keepaspectratio]{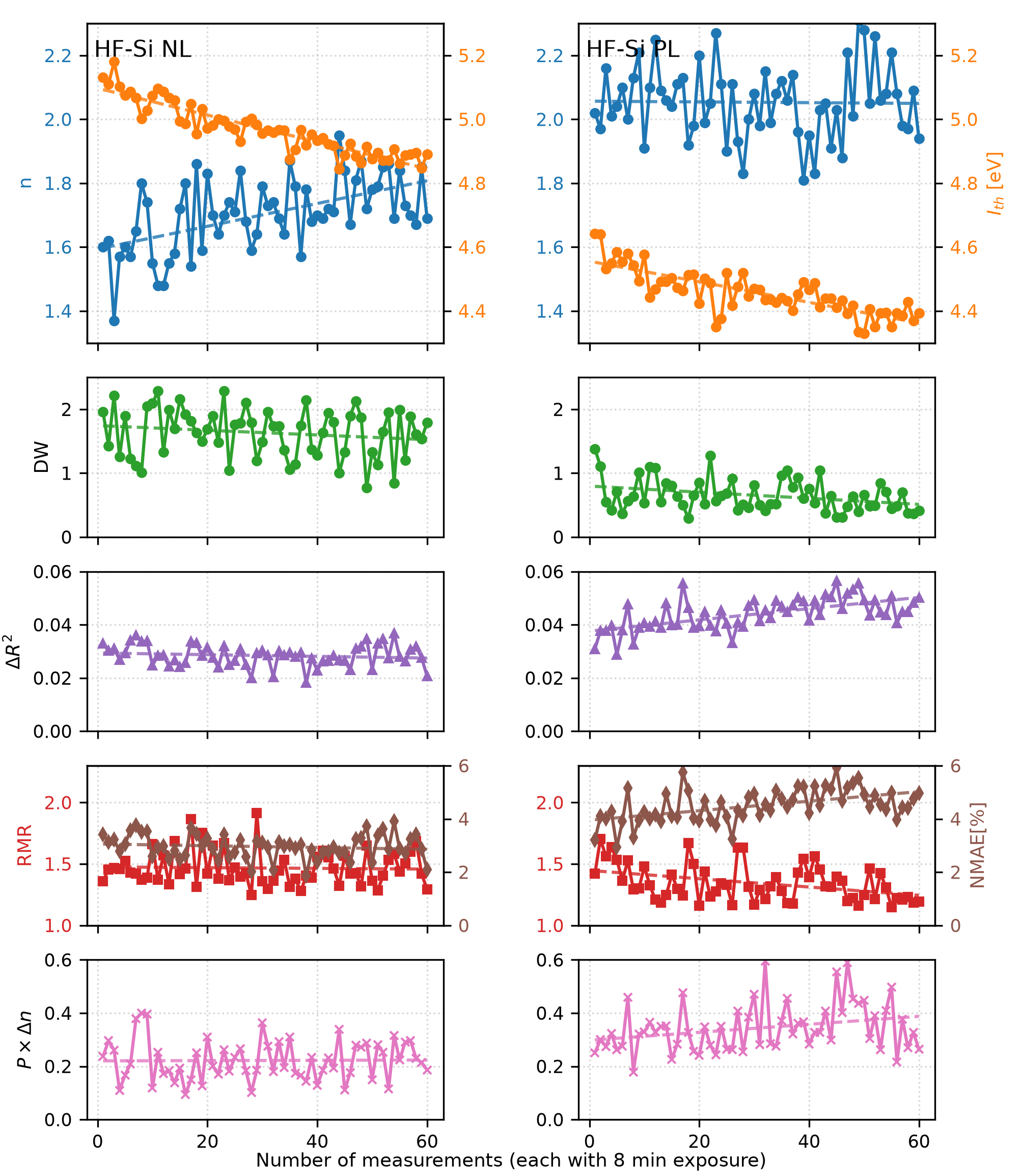}
\caption{Diagnosis of 60 continuous measurements of the HF-treated NL (left) and PL (right) samples. From the top: $n$ (blue, left axis) and $I_{th}$ (orange, right axis); DW; $\Delta R^2$; RMR (red squares, left axis) and NMAE (brown diamonds, right axis); $P \times \Delta n$. Dashed lines are linear trends. The threshold decreases with time for both samples. $n$ of PL is almost constant, and $n$ of NL increases. The means of NMAE and RMR are within the criteria for both samples (Table~\ref{tab:means}). The linear trend of DW (dashed) decreases from 1.74 to 1.52 for NL and stays above $d_L$; for PL, DW decreases to below 0.5 and $\Delta R^2$ increases.}
\label{fig:series}
\end{figure}

Figure~\ref{fig:series} shows the 60 continuous measurements of NL and PL over about 8~h, and Table~\ref{tab:means} lists the mean values of the metrics; each measurement was analyzed as in Fig.~\ref{fig:example}. For NL (Fig.~\ref{fig:series}, left), $n$ increased gradually from about 1.6 to about 1.8 during the series. NMAE was within its criterion in all 60 measurements, and RMR exceeded 1.7 in 3 (at most 1.92); $P \times \Delta n$ was at or below 0.3 in 52 of the 60 measurements (mean 0.22). The linear trend of DW decreased from 1.74 to 1.52 but stayed above $d_L$ (mean 1.63). Individual values fell below $d_L$ in 20 of the 60 measurements, whereas the moving median remained above $d_L$ throughout the series. The change of NL during air exposure was therefore consistent with a single component.

For PL (Fig.~\ref{fig:series}, right), $n$ was nearly constant at about 2.0, whereas $P \times \Delta n$ increased with exposure and exceeded 0.3 in 37 of the 60 measurements (0.18--0.59, mean 0.34). The linear trend of DW decreased from 0.79 to 0.51, and all 60 values as well as the moving median were below $d_L$. The linear trend of $\Delta R^2$ increased from 0.038 to 0.050, and 10 of the 60 values exceeded 0.05 (mean 0.044). As for NL, the means of NMAE and RMR were within the criteria. NMAE exceeded 5\% in 13 of the 60 measurements (at most 5.9\%), and RMR exceeded 1.7 in one. In all of the remaining 46 measurements, which passed the data-quality check, DW was below $d_L$. These spectra therefore fall into the second row of Table~\ref{tab:decision}: a breakdown of the single power-law approximation.

For the lightly doped PH and NH (Supplementary Material S4, Fig.~S5), DW was below $d_L$ in 51 and 48 of the 60 measurements, respectively, from the start of the series (mean 1.15 and 1.17), and the moving median was below $d_L$ throughout. The means of NMAE (3.2--3.3\%) and RMR (1.56--1.61) were within the criteria; RMR exceeded 1.7 in 9 and 13 measurements (above 2.0 in 2 and 3), and among the measurements that passed the data-quality check, DW was below $d_L$ in 42 of 51 and 37 of 47. For these samples, the model-adequacy check was thus flagged from the start of the series, without the progressive decrease observed for PL. Their spectra are consistent with the overlap of the indirect and direct transitions of the Si valence band~\cite{Gobeli1962,Miyazaki1999}.

The mean values in Table~\ref{tab:means} show which metric separates PL from NL and Au. NMAE, RMR, and $\Delta R^2$ of PL are comparable to those of Au and NL, whereas the mean DW of PL (0.65) is far below $d_L$. The residuals of PL deviate smoothly rather than at isolated points, so RMSE and MAE increase together; this keeps RMR low (1.34) and limits the increase in $\Delta R^2$. DW responds to this serial structure, and a passed data-quality check combined with a flagged model-adequacy check indicates a model mismatch.

\subsection{Verification by a two-component fit}

To test whether the flag for PL reflects an overlap of emission components, we fitted the 30th spectrum of PL with a two-component model, a linear sum of two power laws with exponents fixed at $n_1 = 1$ (direct transition) and $n_2 = 2.5$ (indirect transition), by minimizing the MAE in the original space over the same 49 points. For comparison, the single power law with the exponent from the $1/n$-Scan ($n = 2.08$) was fitted in the same way. The exponents were fixed to avoid local minima in the nonlinear fit. The two-component model has $k = 5$ structural parameters (the amplitudes and thresholds of the two components and the background), and the single power-law model has $k = 3$.

\begin{figure}[p]
\centering
\includegraphics[width=\linewidth,height=0.80\textheight,keepaspectratio]{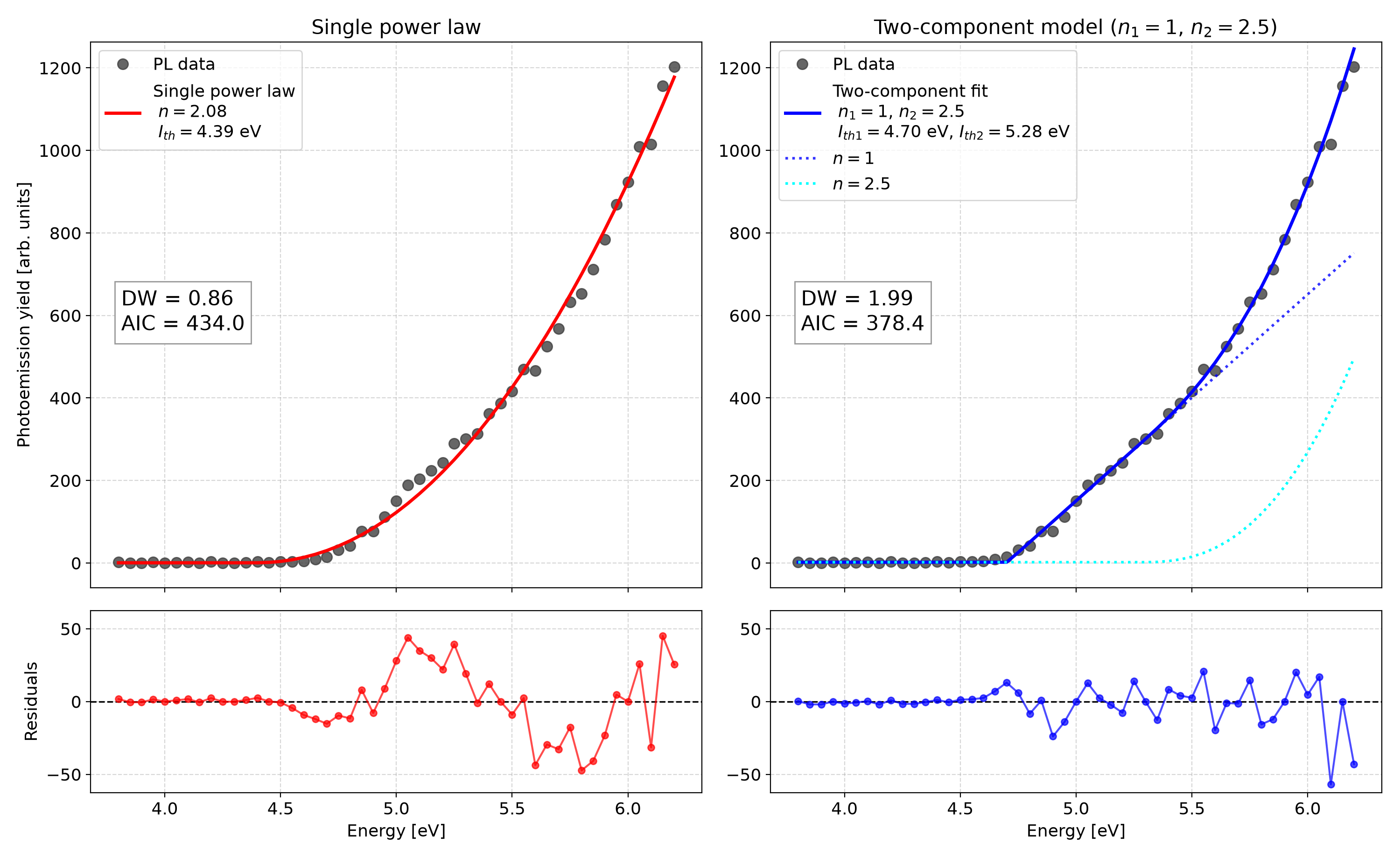}
\caption{Two-component fit of the 30th spectrum of PL with fixed exponents ($n_1 = 1.0$ and $n_2 = 2.5$) and no constraint on the order of the thresholds (right), compared with the single power law at the exponent of the $1/n$-Scan ($n = 2.08$, left), and the residuals. The $n_1 = 1.0$ component sets in at 4.70~eV and the $n_2 = 2.5$ component at 5.28~eV. The AIC (Eq.~(\ref{eq:aic})) improves from 434.0 to 378.4, and the undulation of the residuals is removed (DW from 0.86 to 1.99).}
\label{fig:twocomp}
\end{figure}

The AIC (Eq.~(\ref{eq:aic})) decreased from 434.0 for the single power law to 378.4 for the two-component model. This difference of 56 is far above the level of $\Delta\mathrm{AIC} > 10$, beyond which the weaker model has essentially no empirical support~\cite{Burnham2004}; counting the scanned exponent of the single power law as an additional parameter increases the difference to 58. DW increased from 0.86 to 1.99 (Fig.~\ref{fig:twocomp}). In the optimum, the $n = 1$ component sets in at 4.70~eV and the $n = 2.5$ component at 5.28~eV. When the thresholds were constrained to the order of a clean Si surface, in which the indirect component sets in first~\cite{Gobeli1962}, the optimization placed the $n = 2.5$ component at 4.10~eV and the $n = 1$ component at 5.85~eV, so that the model effectively reduced to a single component; the AIC improved by less than 1, and DW was 1.08 (Supplementary Material S5, Fig.~S10). Because the constrained and free models have the same number of parameters, the improvement of the free model results from the order of the thresholds: the rise from 4.70~eV is described by a component with a slow onset below the threshold of the indirect transition. Within the fixed exponents used here, the direct transition from the valence band near the surface is therefore not supported as the origin of the low-threshold component. The spectrum is consistent with two possible origins that the present data do not distinguish: the valence band on the bulk side, raised toward the Fermi level by downward band bending, and surface or gap states near the Fermi level (Supplementary Material S4 and S5).

\subsection{Relation to the work function}

The work functions $\Phi$ of PH, NH, and NL measured by CPD agreed within 0.1~eV (4.11--4.22~eV), and PL had the largest $\Phi$, as expected from its doping~\cite{Sze2006}. The XPS Si~2p binding energies of the four HF-treated samples agreed within 0.11~eV, suggesting that the Fermi level is pinned at a common position at the surface (Supplementary Material S2). At the start of the series, the PYS threshold was 5.07--5.14~eV for PH, NH, and NL and 4.60~eV for PL (Fig.~S2); the order of $\Phi$ and $I_{th}$ is thus reversed for PL. For PH, NH, and NL, $I_{th}$ exceeds $\Phi$ by 0.9--1.0~eV, consistent with emission from the valence band below the pinned Fermi level. For PL, $I_{th} \approx \Phi$, which suggests emission from occupied states near the Fermi level, consistent with the low-threshold component in Section~4.3. During the series, $I_{th}$ of PH, NH, and NL decreased by 0.27--0.32~eV up to the CPD measurement after the series, and $\Phi$ decreased by 0.17--0.28~eV; this agreement within about 0.1~eV is consistent with the threshold following the vacuum level (Table~S4).

\section{Discussion}

The diagnostics separate the causes of a poor single power-law fit by combining metrics that respond to different properties of the residuals: NMAE and RMR respond to the amount of noise and to isolated anomalies, whereas DW responds to the serial structure produced by a smooth deviation of the model. For PL, the means of NMAE, RMR, and $\Delta R^2$ were comparable to those of Au, so a single goodness-of-fit value would not have distinguished the overlap of components from noisy data; only DW, together with the increase in $P \times \Delta n$, indicated the mismatch. The framework does not identify the origin of the components. Instead, it identifies the spectra that require a multi-component analysis, and the subsequent analysis, here a two-component fit with fixed exponents, gives the order of the thresholds.

The DW criterion $d_L$ was derived for linear least-squares regression with Gaussian errors, whereas the present fit is nonlinear and minimizes the MAE. We therefore use DW as an indicator rather than as a formal test and rely on the moving-median rule for the reliability of the check. With $N = 49$, a weak overlap of components may not lower the moving median (Supplementary Material S4 and S6). The criteria for NMAE and RMR depend on the measurement conditions, such as the energy range, the step, and the counting time, which determine the noise level and the fraction of baseline points among the $N$ points. For other instruments or conditions, these criteria should be set from a single-component reference measured under the same conditions, as was done with Au in this work. The criterion $P \times \Delta n \le 0.3$ follows from the spacing of Kane's exponents; for organic semiconductors, for which exponents of 2.5 and 3 have been discussed~\cite{Grzibovskis2018,Grzibovskis2016,Kochi1970}, and for materials with a continuous distribution of exponents, the criterion should be set from the exponent spacing of the target system. The estimation assumes a low background. The RIKEN KEIKI AC series keeps $Y_{bg}$ at a few counts per second over the measurement range~\cite{Yagyu2022}, and simulated spectra with $Y_{bg} = 2$ showed an apparent increase in $n$ of 0.02--0.05, which is small compared with the spacing of 0.5 and with the changes of 0.1--0.3 observed in the series. For instruments with a high or energy-dependent background, the background should be removed before the analysis, because a large background raises the estimated $I_{th}$~\cite{Yagyu2025} and, through the nonlinearity of the $1/n$ transformation, also $n$.

The framework reports the three checks together with their values. How these checks are weighted in choosing the next measurement, and how the diagnosis is fed back in a closed loop, are design decisions on the decision-making side of an autonomous experimental system and are beyond the scope of this work.

\section{Conclusions}

We have constructed a residual-diagnosis framework for the single power-law analysis of PYS spectra. The framework evaluates the uncertainty of the effective exponent, scanned continuously by the $1/n$-Scan, with the perplexity of the Akaike weights; the data quality with NMAE and RMR; and the adequacy of the single power-law model with DW. It reports the three checks in parallel. With criteria set from an Au reference, the framework found the change of heavily doped n-type Si during air exposure to be consistent with a single component, flagged the model adequacy of lightly doped p- and n-type Si from the start of the series, and detected the breakdown of the single power-law approximation in heavily doped p-type Si while the means of the data-quality metrics remained within their criteria. A two-component fit confirmed the breakdown for the p-type sample, with components setting in at 4.70 and 5.28~eV. The framework thus identifies the spectra that require a multi-component analysis without assuming an emission mechanism.

\section*{Acknowledgments}
The authors thank Yubin Liu for the PYS and CPD measurements of the Si and Au samples.

\section*{Funding}
This work was supported by the Research Center for Electronic and Optical Materials in the National Institute for Materials Science (NIMS) and the JST-Mirai Program ``Development of Materials Design Workflow and Data Library for Materials Foundry,'' Grant No.~JPMJMI18G5.

\section*{Declaration of AI use}
During the preparation of this work, the authors used Claude (Anthropic; Claude Sonnet 5 and Claude Opus 5.5) in order to improve the English language and readability, and to generate analysis code. After using these tools, the authors reviewed and edited the content as needed and take full responsibility for the content of the publication.

\section*{Disclosure statement}
No potential conflict of interest was reported by the author(s).

\section*{Author contributions}
Shinjiro Yagyu: Conceptualization, Methodology, Software, Formal analysis, Investigation, Writing -- original draft. Takahiro Nagata: Investigation, Resources, Writing -- review \& editing. Yoshiyuki Nakajima: Resources, Writing -- review \& editing.

\section*{Data availability statement}
The analysis source code (Python implementation of the framework) is openly available on GitHub at https://github.com/s-yagyu/n-pys, together with the raw PYS spectra of the Au measurements and of measurements 1--60 of the HF-treated Si series (Figs.~2--4, S3--S5) as sample data. The remaining data that support the findings of this study will be openly available in the NIMS Materials Data Repository (MDR; https://mdr.nims.go.jp/) upon publication, together with the data bundled on GitHub.

\clearpage

\clearpage

\clearpage

\clearpage

\clearpage
\begin{table}[p]
\centering
\caption{Interpretation of the combinations of the checks. A flagged estimation check ($P \times \Delta n > 0.3$) indicates a wide range of supported $n$ in any row.}
\label{tab:decision}
\small
\begin{tabular}{L{2.7cm}L{2.8cm}L{4.3cm}L{4.1cm}}
\toprule
Data-quality check (NMAE, RMR) & Model-adequacy check (DW) & Interpretation & Action \\
\midrule
Passed & Passed (DW $> d_L$) & A single power law describes the spectrum & Accept $n$ and $I_{th}$ (with $P \times \Delta n$ as the width) \\
Passed & Flagged (DW $< d_L$) & Breakdown of the single power law (e.g., overlap of components) & Multi-component analysis \\
Flagged & Not interpreted & Excessive noise or isolated anomalies & Repeat the measurement; remove spikes \\
\bottomrule
\end{tabular}

\vspace{12mm}
\caption{Mean values of the diagnostic metrics for the Au reference (10 measurements, used to set the criteria) and the HF-treated Si NL and PL samples (60 measurements each), and the criteria.}
\label{tab:means}
\begin{tabular}{lcccl}
\toprule
Metric & Au & Si NL & Si PL & Criterion \\
\midrule
$P \times \Delta n$ & 0.27 & 0.22 & 0.34 & $\le 0.3$ (upper warning 0.5) \\
NMAE (\%) & 4.5 & 3.0 & 4.5 & $\le 5.0$ \\
RMR & 1.62 & 1.46 & 1.34 & $\le 1.7$ (upper warning 2.0) \\
DW & 2.15 & 1.63$^{a}$ & 0.65 & $> 1.46$ ($d_L$) \\
$\Delta R^2$ & 0.042 & 0.028 & 0.044 & $\le 0.05$ \\
\bottomrule
\end{tabular}

\vspace{1mm}
{\footnotesize $^{a}$Linear trend from 1.74 to 1.52; 20 of the 60 individual values are below $d_L$.}
\end{table}

\clearpage
\setcounter{section}{0}\setcounter{figure}{0}\setcounter{table}{0}\setcounter{equation}{0}
\renewcommand{\thefigure}{S\arabic{figure}}
\renewcommand{\thetable}{S\arabic{table}}
\renewcommand{\theequation}{S\arabic{equation}}
\renewcommand{\thesection}{S\arabic{section}}
\renewcommand{\thesubsection}{S\arabic{section}.\arabic{subsection}}
\renewcommand{\theHfigure}{S\arabic{figure}}
\renewcommand{\theHtable}{S\arabic{table}}
\renewcommand{\theHequation}{S\arabic{equation}}
\renewcommand{\theHsection}{S\arabic{section}}
\renewcommand{\theHsubsection}{S\arabic{section}.\arabic{subsection}}
\setlength{\belowcaptionskip}{3pt}
\makeatletter
\renewcommand{\@biblabel}[1]{[S#1]}
\def\@bibitem#1{\item\if@filesw\immediate\write\@auxout{\string\bibcite{#1}{S\the\value{\@listctr}}}\fi\hyper@anchorstart{cite.#1}\relax\hyper@anchorend\ignorespaces}
\makeatother

\begin{center}
{\large\bfseries Supplementary Material for\par
Residual diagnostics for detecting the breakdown of the single power-law approximation in photoelectron yield spectroscopy: application to doped silicon\par}
\vspace{4mm}
Shinjiro Yagyu, Takahiro Nagata, and Yoshiyuki Nakajima\par
\end{center}

\noindent Section, equation, figure, and table numbers without the prefix ``S'' refer to the main text.

\section{Samples and measurement methods}

The samples are four Si(100) wafers: p-type high resistivity (PH), p-type low resistivity (PL), n-type high resistivity (NH), and n-type low resistivity (NL). From Irvin's curves~\cite{S-Irvin1968}, the impurity concentration of the low-resistivity samples (PL and NL, $<0.02~\Omega$\,cm) is on the order of $10^{18}$--$10^{19}$~cm$^{-3}$, and these samples are expected to be degenerate.

Each sample was cleaned ultrasonically in acetone and ethanol, immersed in HF solution for 2~min, and rinsed with pure water to terminate the surface with hydrogen. To limit oxidation in air, all CPD and PYS measurements were started within 3~h of the HF treatment. After the HF treatment, rinsing, and drying, the CPD was measured first, followed by the continuous PYS measurements.

XPS spectra were recorded with a ULVAC-PHI Quantera with monochromated Al K$\alpha$ radiation (1486.6~eV); the pass energy was 140~eV for the wide scan and 55~eV for the narrow scans. The binding energy was calibrated by setting the C~1s peak of contaminant carbon to 284.9~eV. The XPS samples were HF-treated separately in the same way and introduced into the vacuum chamber immediately.

The CPD was measured in air with a RIKEN KEIKI FAC-2 Kelvin probe. The work function of the probe was calibrated against the PYS threshold of a rolled Au plate (99.999\%) stored in air for a long time with a stable surface potential. The repeatability was a few mV or less; remounting the sample gave an error of about $\pm$50~mV, caused by small differences (tens of $\mu$m) in the distance between the sample surface and the probe.

PYS spectra were measured in air with a RIKEN KEIKI AC-5 from 3.8 to 6.2~eV in 0.05~eV steps (49 points, about 8~min per measurement) for all samples, including the Au reference. The laboratory temperature was $23 \pm 2~^{\circ}$C. The PYS signal integrates the emission from within the escape depth of the photoelectrons, a few nanometers (about 2.5~nm for Si~\cite{S-Miyazaki1999}), and therefore reflects band bending within this depth as well as the outermost surface.

The three methods provide complementary quantities (Table~\ref{S-tab:methods}). XPS gives the binding energy $BE = (E_F - E_{VBM}) + (E_{VBM} - E_{core})$; because the second term is a constant of Si, $\Delta BE = \Delta(E_F - E_{VBM})$ gives the relative position of the Fermi level at the surface. The CPD gives the work function $\Phi = E_{vac} - E_F$, which includes the shift of the vacuum level by the surface dipole. PYS gives the threshold $I_{th}$ and the exponent $n$. When $I_{th}$ is the threshold of emission from the valence-band maximum (VBM),
\begin{equation}
I_{th}\,(\mathrm{PYS}) = \Phi\,(\mathrm{CPD}) + (E_F - E_{VBM}),
\end{equation}
and when several components overlap, the single power-law $I_{th}$ is an effective value (Section~\ref{S-sec:band}).

\begin{table}[H]
\centering
\caption{Measurement methods and their results for the HF-treated Si surfaces.}
\label{S-tab:methods}
\begin{tabular}{p{1.4cm}p{3.6cm}p{3.0cm}p{6.2cm}}
\toprule
Method & Quantity & Depth & Result for the HF-treated Si surfaces \\
\midrule
XPS & Binding energy $BE$ (Si~2p) & A few nm & Si~2p within 0.11~eV for the four samples: common surface Fermi-level position \\
CPD & Work function $\Phi$ & Outermost surface & PH $\approx$ NH $\approx$ NL within 0.1~eV; PL largest and decreasing under PYS \\
PYS & Threshold $I_{th}$, exponent $n$ & Escape depth (a few nm) & PH, NH, NL: valence band near the surface ($I_{th} - \Phi \approx 0.9$--1.0~eV, $n \approx 1.5$); PL: occupied states near $E_F$ ($I_{th} \approx \Phi$, $n \approx 2.0$) \\
\bottomrule
\end{tabular}
\end{table}

\section{XPS and CPD of the HF-treated surfaces}

\begin{figure}[p]
\centering
\includegraphics[width=0.48\linewidth]{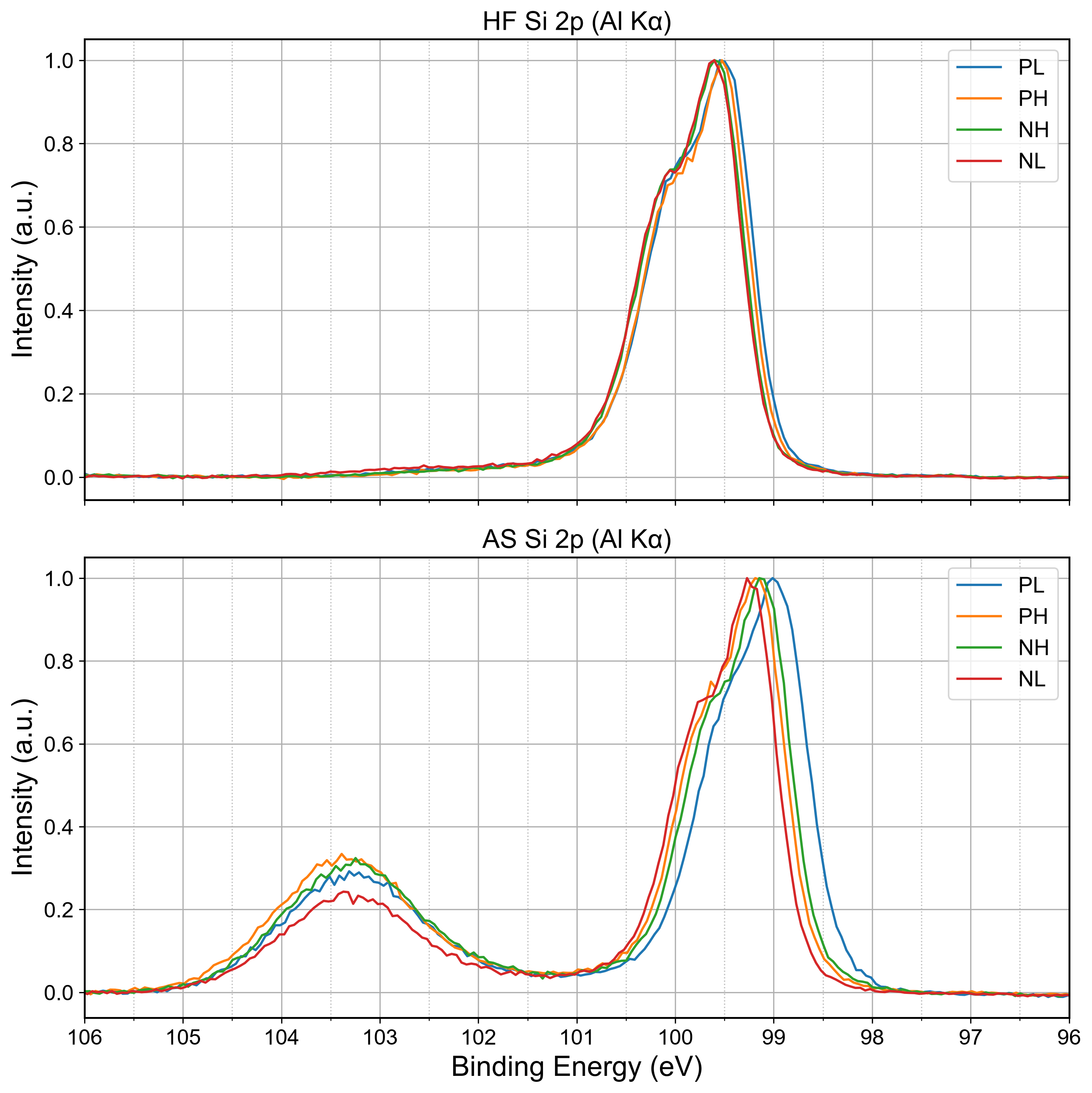}\hfill
\includegraphics[width=0.48\linewidth]{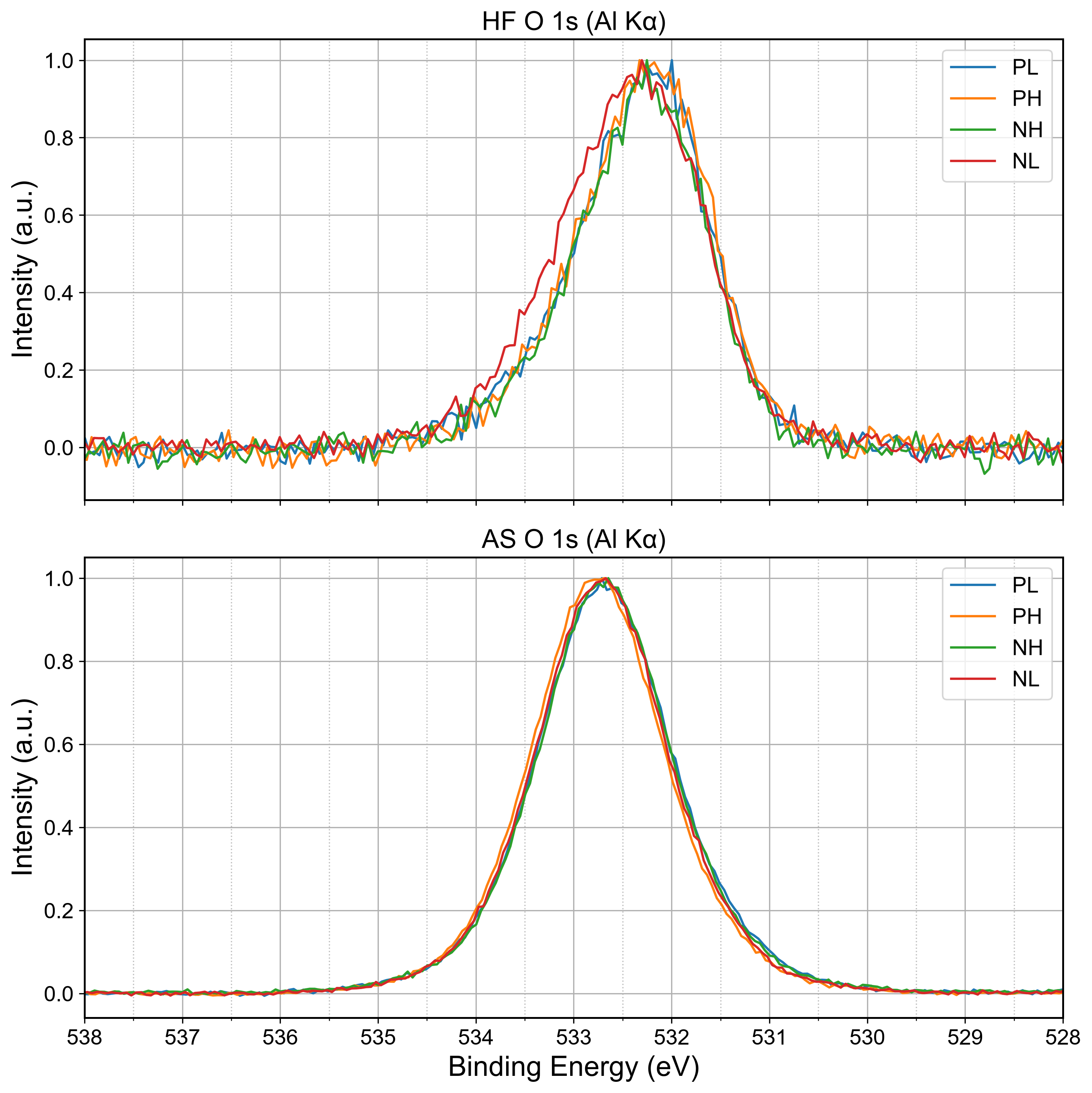}
\caption{Si~2p and O~1s XPS spectra of the HF-treated and AS surfaces.}
\label{S-fig:xps}
\end{figure}

The Si~2p and O~1s spectra of the HF-treated and as-received (AS) surfaces are shown in Fig.~\ref{S-fig:xps}. On the HF-treated surface, a weak O~1s peak and a shoulder on the high-binding-energy side of Si~2p indicate initial oxidation or an adsorbed layer. Such a layer can affect PYS and CPD in two ways. A change of the surface dipole shifts the vacuum level and changes $\Phi$ and $I_{th}$ by the same amount. A disordered layer scatters the photoelectrons; scattering reduces the unscattered direct-transition emission ($n = 1$) and adds to the emission at lower energy~\cite{S-Gobeli1962}, which raises the effective exponent. The O~1s intensity after the HF treatment is much weaker than that of the AS samples, and the exponents at the start of the measurements (NL: $n \approx 1.5$; PL: $n \approx 2.0$) differ with the doping.

Table~\ref{S-tab:xps} lists the Si~2p binding energies. The HF treatment shifted them by $+0.35$ to $+0.53$~eV. On the HF-treated surfaces, the binding energies of the four samples lie within 0.11~eV, whereas the bulk Fermi levels of PL and NL differ by about 1.1~eV. This suggests that the Fermi level at the surface is pinned at a common position and that the band bending is set by the bulk Fermi level. Because the binding energy is referenced to the C~1s peak, only relative positions are used.

\begin{table}[H]
\centering
\caption{XPS Si~2p binding energy (eV) of the AS and HF-treated surfaces.}
\label{S-tab:xps}
\begin{tabular}{lccc}
\toprule
Sample & AS & HF & $\Delta$(HF$-$AS) \\
\midrule
PL & 98.88 & 99.41 & $+0.53$ \\
PH & 99.10 & 99.45 & $+0.35$ \\
NH & 99.04 & 99.50 & $+0.46$ \\
NL & 99.17 & 99.52 & $+0.35$ \\
\bottomrule
\end{tabular}
\end{table}

Table~\ref{S-tab:cpd} lists the work functions measured by CPD before and after the PYS measurements. The absolute scale is set by the Au reference of each series; differences between samples within a series and changes of one sample within a series do not depend on it. On the HF-treated surfaces, PH, NH, and NL agree within about 0.1~eV in every series (4.11--4.22~eV), consistent with the common pinning position obtained by XPS. PL has the largest $\Phi$, which varies between series (4.62--4.80~eV) and decreases by 0.36~eV after a single PYS measurement (series n1) and by 0.37~eV after five (series n2); $\Phi$ of NL decreases by 0.04 and 0.15~eV, respectively. The PL surface is therefore modified by the PYS measurement in air (ultraviolet irradiation), and $\Phi$ of PL during the PYS measurement is lower than the value measured before it.

\begin{table}[H]
\centering
\caption{Work function (eV) measured by CPD. ``After PYS'' is measured after the number of PYS measurements in the third column. Au ref.\ is the value assigned to the Au reference in each series. CPD1 and CPD2 are the series plotted in Fig.~\ref{S-fig:cpd}; series n5 (CPD1) brackets the continuous PYS measurements of Fig.~3 and Fig.~\ref{S-fig:nhph}.}
\label{S-tab:cpd}
\small
\begin{tabular}{lllcccccc}
\toprule
Series & State & PYS & PL & PH & NH & NL & Au ref. \\
\midrule
n1 (CPD2) & HF & 0 & 4.76 & 4.22 & 4.20 & 4.11 & 4.88 \\
n1 (CPD2) & HF, after PYS & 1 & 4.40 & 4.18 & 4.16 & 4.07 & 4.88 \\
n2 & HF & 0 & 4.62 & 4.22 & 4.22 & 4.18 & 4.88 \\
n2 & HF, after PYS & 5 & 4.25 & 4.08 & 4.07 & 4.03 & 4.88 \\
n5 (CPD1) & HF & 0 & 4.80 & 4.18 & 4.19 & 4.11 & 4.88 \\
n5 (CPD1) & HF, after PYS & 99$^{a}$ & 4.02 & 3.96 & 3.91 & 3.94 & 4.88 \\
n3 & AS & 0 & 4.52 & 4.22 & 4.28 & 4.22 & 4.90 \\
n3 & AS, after PYS & 20 & 4.44 & 4.46 & 4.26 & 4.20 & 4.90 \\
n4 & AS & 0 & 4.50 & 4.33 & 4.37 & 4.25 & 4.90 \\
n4 & AS, after PYS & 3 & 4.46 & 4.26 & 4.30 & 4.20 & 4.90 \\
\bottomrule
\end{tabular}

\vspace{1mm}
{\footnotesize $^{a}$61 for NL.}
\end{table}

\begin{figure}[p]
\centering
\includegraphics[width=\linewidth,height=0.6\textheight,keepaspectratio]{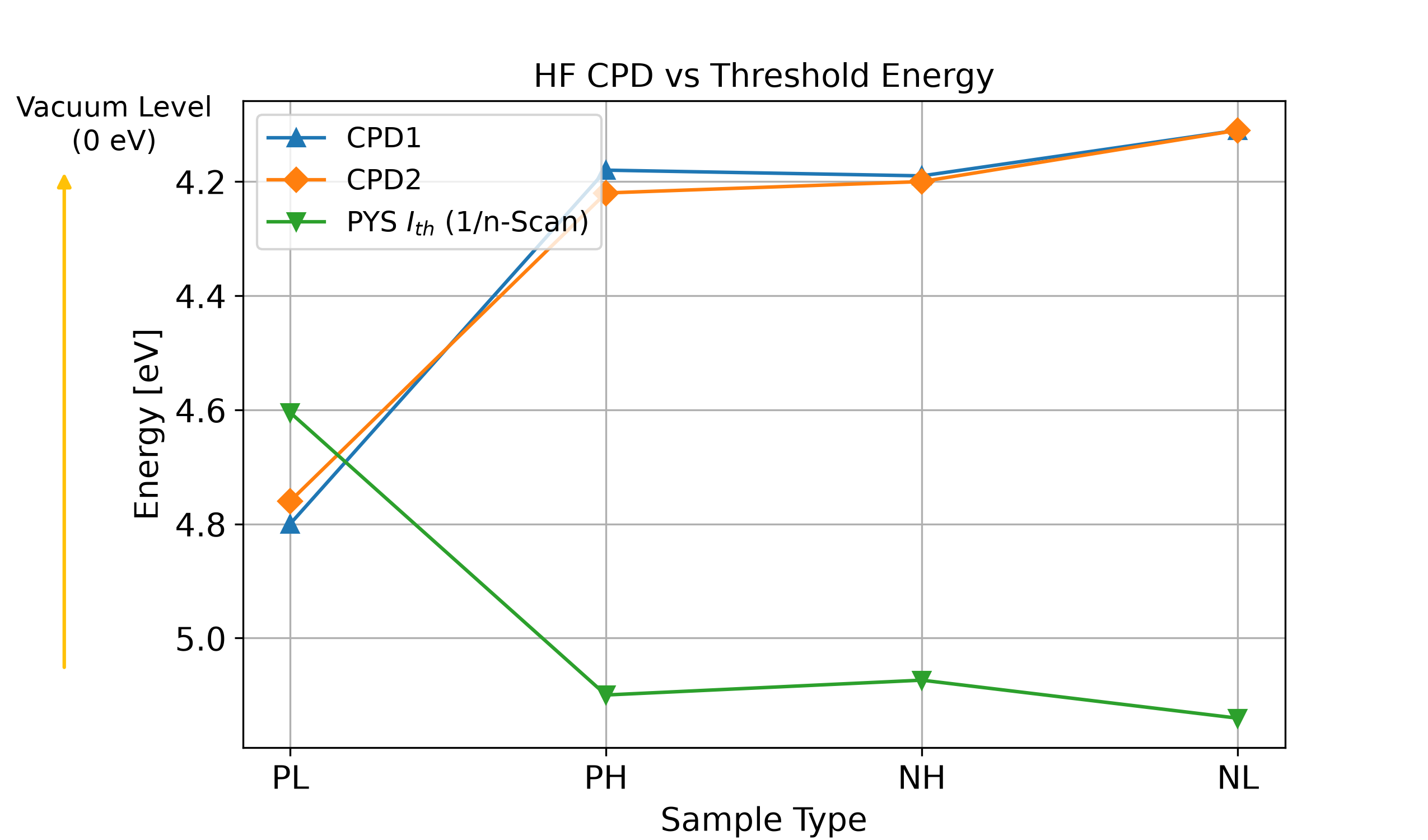}
\caption{Work function from CPD (series CPD1 and CPD2 of Table~\ref{S-tab:cpd}) and PYS threshold $I_{th}$ (mean of measurements 1--3, Table~\ref{S-tab:phiith}) of the HF-treated samples. The energy is referenced to the vacuum level and increases downward. The order of $\Phi$ (PL $>$ PH $\approx$ NH $>$ NL) and that of $I_{th}$ (NL $>$ NH $\approx$ PH $>$ PL) are reversed.}
\label{S-fig:cpd}
\end{figure}

Table~\ref{S-tab:phiith} compares $\Phi$ of series n5 with the PYS thresholds at the start and at the end of the continuous measurements, and Fig.~\ref{S-fig:cpd} compares $\Phi$ of two series with the thresholds at the start. For PH, NH, and NL, $I_{th} = 5.07$--5.14~eV at the start, and $I_{th} - \Phi$ is 0.88--1.03~eV at the start and 0.82--0.93~eV at the end. $I_{th} - \Phi$ equals $E_F - E_{VBM}$ at the surface when $I_{th}$ is the threshold of the VBM; because the single power-law $I_{th}$ can lie above the indirect-transition threshold when the indirect and direct transitions overlap (Section~\ref{S-sec:band}), these values are upper estimates of $E_F - E_{VBM}$. The decrease of $I_{th}$ up to the CPD measurement after the series (0.27--0.32~eV) agrees with the decrease of $\Phi$ (0.17--0.28~eV) within about 0.1~eV, which is consistent with the threshold following the vacuum level and with a change of $E_F - E_{VBM}$ at the surface of 0.1~eV or less. For PL, $I_{th}$ of the first measurements (4.60~eV) lies between $\Phi$ before the series (4.80~eV) and $\Phi$ after a single PYS measurement in series n1 (4.40~eV); within the accuracy of these values, the threshold of PL is at the Fermi level.

\begin{table}[H]
\centering
\caption{Work function (series n5) and PYS threshold (eV) at the start (measurements 1--3) and at the end of the continuous measurements. The end values are measurements 97--99 for PH and NH, whose CPD was measured after 99 measurements, and measurements 58--60 for NL (CPD after 61 measurements) and PL (CPD after 99 measurements; the spectra analyzed are measurements 1--60). For PL, $\Delta\Phi$ and $I_{th} - \Phi$ at the end therefore include the change during measurements 61--99.}
\label{S-tab:phiith}
\small
\begin{tabular}{lcccccccc}
\toprule
Sample & $\Phi$ start & $\Phi$ end & $I_{th}$ start & $I_{th}$ end & $\Delta\Phi$ & $\Delta I_{th}$ & $I_{th}-\Phi$ start & $I_{th}-\Phi$ end \\
\midrule
PL & 4.80 & 4.02 & 4.60 & 4.40 & $-0.78$ & $-0.21$ & $-0.20$ & 0.38 \\
PH & 4.18 & 3.96 & 5.10 & 4.78 & $-0.22$ & $-0.32$ & 0.92 & 0.82 \\
NH & 4.19 & 3.91 & 5.07 & 4.78 & $-0.28$ & $-0.29$ & 0.88 & 0.87 \\
NL & 4.11 & 3.94 & 5.14 & 4.87 & $-0.17$ & $-0.27$ & 1.03 & 0.93 \\
\bottomrule
\end{tabular}
\end{table}

\section{Au reference}

The proposed procedure was first applied to a single-component reference, a rolled Au plate (99.999\%), which is little affected by band bending and degradation in air. Figure~\ref{S-fig:autrend} shows the 10 continuous measurements, and Fig.~\ref{S-fig:au4} the detailed output for the fourth measurement.

Throughout the series, $n$ was 1.65--2.02 (mean 1.84), slightly below the Fowler value for metals ($n = 2.0$), and is treated as an effective exponent. NMAE (3.2--6.5\%, mean 4.5\%) was higher than for the HF-treated Si samples (means 3.0--4.5\%). DW was 1.78--3.19 (mean 2.15, median 2.06) and lay within $d_L < \mathrm{DW} < 4 - d_L$ (1.46--2.54) except for one measurement (No.~7, DW $= 3.19$); this isolated value is treated as a fluctuation by the moving-median rule (Section~2.4). RMR was 1.50--1.75 (mean 1.62), and $\Delta R^2$ was 0.030--0.058 (mean 0.042). Because the criteria were set from these means, NMAE exceeded 5\% in measurements 1 and 4 (6.5\% and 5.8\%), and RMR exceeded 1.7 in measurements 3 and 7 (1.71 and 1.75). The residuals are close to white noise without structural undulation, and a single power law describes the Au spectra within the noise.

\section{Doped Si: results and emission processes}

\subsection{Continuous measurements of NH and PH}

Figure~\ref{S-fig:nhph} shows the 60 continuous measurements of the HF-treated NH and PH (NL and PL are shown in Fig.~3). With the exposure to air and to the PYS measurements, $I_{th}$ decreased by 0.20--0.21~eV during the 60 measurements shown and by 0.29--0.32~eV up to the 99th measurement, after which $\Phi$ had decreased by 0.22--0.28~eV (Table~\ref{S-tab:phiith}); the decrease is attributed to a shift of the vacuum level by the change of the surface dipole. The exponent increased from 1.44--1.45 to 1.54--1.58, as for NL (Section~\ref{S-sec:nl}). DW of NH and PH was below $d_L$ in 48 and 51 of the 60 measurements, respectively, from the start of the series (mean 1.17 and 1.15), while the means of NMAE (3.2--3.3\%) and RMR (1.56--1.61) were within the criteria. RMR exceeded 1.7 in 13 measurements of NH and 9 of PH (above 2.0 in 3 and 2; at most 2.27); these isolated anomalies cause the data-quality check to be flagged for the measurements concerned. Among the measurements that passed the data-quality check, DW was below $d_L$ in 37 of 47 (NH) and 42 of 51 (PH): the spectra deviate from a single power law throughout the series, and $n \approx 1.5$ is an effective value. The origin of this deviation is discussed in Section~\ref{S-sec:band}.

\subsection{Band bending and emission processes}
\label{S-sec:band}

The band diagrams constructed from the XPS, CPD, and PYS results (Tables~\ref{S-tab:xps}--\ref{S-tab:phiith}) are shown in Fig.~\ref{S-fig:band}. The Si~2p binding energies and the work functions suggest that the Fermi level at the HF-treated surface is pinned at a common position. With the surface Fermi level fixed, the band bending is set by the bulk Fermi level: slightly upward for NL, nearly flat for NH, and downward for PH and PL. From $I_{th} - \Phi$ (Table~\ref{S-tab:phiith}), the pinning position is at most 0.9--1.0~eV above the VBM at the surface; the absolute value carries the uncertainty of the Au reference of the CPD and the dependence of $I_{th}$ on the spectral model. A similar position, 0.85~eV above the VBM, was reported for the quasi-Fermi level of p-type Si covered with an ultrathin oxide during PYS measurements~\cite{S-Miyazaki1999}.

For PH, NH, and NL, the threshold ($I_{th} \approx 5.1$~eV) and $I_{th} - \Phi$ suggest emission from the valence band near the surface. For the lightly doped PH and NH, the depletion width (on the order of 1~$\mu$m) is much larger than the escape depth of the photoelectrons, so that the bands are flat within the escape depth; for NL, the upward bending places the highest VBM at the surface. The valence-band emission of Si consists of an indirect transition from the VBM (Kane's $n = 5/2$) and a direct transition ($n = 1$) with a threshold about 0.3~eV higher (5.15 and 5.45~eV for cleaved Si)~\cite{S-Gobeli1962}. PYS spectra of Si-based samples are analyzed with these two components, and $n = 3$ is also used for the indirect component~\cite{S-Miyazaki1999}. A single power law fitted to their sum gives an effective exponent between 1 and 5/2, and its threshold can lie above the indirect-transition threshold because the indirect component rises slowly near its onset. The exponent $n \approx 1.5$ is therefore interpreted as the effective exponent of the overlapping indirect and direct transitions. The low DW of PH and NH is consistent with this overlap; for NL, the deviation does not produce a persistent decrease of DW at the noise level of these spectra (Section~\ref{S-sec:nl}). In Kane's theory, $n = 3/2$ is also obtained for specular transitions from volume states at a perfect surface~\cite{S-Kane1962} (Table~\ref{S-tab:kane}); the present spectra do not distinguish this mechanism from the overlap of the two transitions.

Filled surface and gap states give emission below the valence-band threshold, starting near $\Phi \approx 4.2$~eV. Miyazaki et al.\ observed such emission on hydrogen-terminated n$^{+}$-Si(100) and attributed it to slight oxidation during air exposure and rinsing~\cite{S-Miyazaki1999}. Because the density of these states (on the order of $10^{12}$--$10^{13}$~cm$^{-2}$) is small compared with that of the valence band within the escape depth (on the order of $10^{15}$~cm$^{-2}$), the emission is weak, and in the single power-law analysis of PH, NH, and NL it is not separated from the valence-band emission.

For PL, the impurity concentration is on the order of $10^{18}$--$10^{19}$~cm$^{-3}$, and degeneracy and band-gap narrowing~\cite{S-Irvin1968,S-Sze2006} place the Fermi level in the valence band of the bulk. The depletion width decreases to a few to several tens of nanometers, and the downward bending raises the valence band on the bulk side toward the Fermi level; this corresponds to the volume effect of Gobeli and Allen, in which band bending changes the position of the valence band within the escape depth and shifts the effective threshold~\cite{S-Gobeli1962}. The observed $I_{th} \approx \Phi$ (Table~\ref{S-tab:phiith}) is explained by emission from occupied states near $E_F$, provided either by the valence band on the bulk side raised by the band bending or by surface and gap states near $E_F$ (Section~\ref{S-sec:twocomp}). The spectrum of PL contains this low-threshold emission and the emission from the valence band nearer the surface; the effective exponent $n \approx 2.0$ of the single power-law fit reflects this overlap, which is analyzed with the two-component model in Section~4.3 and Section~\ref{S-sec:twocomp}.

\subsection{Time series of NL}
\label{S-sec:nl}

Figure~\ref{S-fig:nl} shows the change of NL during the series (left: measurements 1--3; right: measurements 58--60). $I_{th}$ decreased by 0.27~eV and $\Phi$ by 0.17~eV, and $I_{th} - \Phi$ stayed at 0.93--1.03~eV (Table~\ref{S-tab:phiith}). The decrease of the threshold is therefore attributed to a lowering of the vacuum level by the change of the surface dipole (adsorption of water and initial oxidation), with the surface Fermi level remaining pinned. The exponent increased from 1.53 (measurements 1--3) to 1.75 (measurements 58--60); PH and NH show a similar increase. Scattering of the photoelectrons removes the unscattered direct-transition emission and adds to the lower-energy emission~\cite{S-Gobeli1962}, which lowers the share of the direct transition and raises the effective exponent of the overlapping transitions; adsorption and initial oxidation disorder the hydrogen-terminated surface, and the increase of the effective exponent is consistent with this increase of scattering. The linear trend of DW decreased from 1.74 to 1.52 and stayed above $d_L$; individual values fell below $d_L$ in 20 of the 60 measurements, while the moving median over ten measurements stayed above $d_L$ throughout the series, and $P \times \Delta n$ exceeded 0.3 in 8 measurements. A two-component fit with the exponents of Section~\ref{S-sec:twocomp} improved the AIC of NL by a median of 9 (more than 10 in 27 of the 60 spectra), compared with 53 for PL; the $n_2 = 2.5$ component set in first (median thresholds 4.72 and 5.12~eV), the order of the indirect and direct transitions of clean Si~\cite{S-Gobeli1962}. The overlap of the two transitions is thus present in NL but small relative to the noise, and at the noise level of these spectra the change is consistent with a single component. The decrease of $\Phi$ of PL after a single PYS measurement (Table~\ref{S-tab:cpd}) suggests that the ultraviolet irradiation during the PYS measurement in air contributes to these surface changes in addition to the exposure time.

\subsection{As-received samples}

For the AS surfaces with a native oxide, the CPD order was PL $>$ NH $>$ PH $\approx$ NL; apart from a small deviation of PH, the order expected from the bulk Fermi levels is preserved. Figures~\ref{S-fig:aspl} and \ref{S-fig:asph} show 20 continuous measurements of the AS surfaces. The order of the PYS threshold was NL $\approx$ NH $\approx$ PH $>$ PL, and, as on the HF-treated surfaces, only PL showed a low value. The exponent of the AS surfaces was mostly 1.6--2.0 (mean 1.72--1.84) and did not show a clear origin (1.5 or 2.0), which suggests that scattering by the oxide dominates.

\section{Two-component model for the PL sample}
\label{S-sec:twocomp}

The 30th spectrum of the continuous measurement of PL was fitted with a two-component model based on the analysis of Gobeli and Allen~\cite{S-Gobeli1962} by minimizing the MAE, and the result was compared with the single power-law model. For the single power-law model, the effective exponent $n = 2.08$ obtained by the $1/n$-Scan was used. In the two-component model, a component with a linear rise ($n_1 = 1.0$, the direct-transition term of Gobeli and Allen) and an indirect-transition component from the valence band ($n_2 = 2.5$) overlap independently; the exponents were fixed to prevent the nonlinear fit with several parameters from being trapped in local minima. Two cases were compared: a free model with no constraint on the order of the thresholds (Fig.~4) and a constrained model in which the indirect transition ($n_2 = 2.5$) sets in first (Fig.~\ref{S-fig:constrained}). The single power-law and two-component models were fitted to the same $N = 49$ points (3.8--6.2~eV), so that the AIC values are directly comparable. The numbers of structural parameters are $k = 3$ for the single power-law model (threshold, amplitude, and background) and $k = 5$ for the two-component model (the thresholds and amplitudes of the two components and the background). The AIC values in this section were computed with Eq.~(3) of the main text.

In the free model, $\mathrm{AIC} = 378.4$, an improvement of $\Delta\mathrm{AIC} \approx 56$ over the single power law ($\mathrm{AIC} = 434.0$), and DW increased from 0.86 to 1.99. Counting the scanned exponent $n = 2.08$ of the single power-law model as an additional parameter gives $\Delta\mathrm{AIC} \approx 58$. In the optimum, $I_{th}(n{=}1.0) = 4.70~\mathrm{eV} < I_{th}(n{=}2.5) = 5.28~\mathrm{eV}$: the component with the linear rise sets in about 0.6~eV below the indirect-transition component, close to the work function of PL (Table~\ref{S-tab:cpd}). Two possible origins are consistent with this threshold. (i) The valence band on the bulk side, raised toward the Fermi level by the downward band bending (Section~\ref{S-sec:band}); the difference of 0.6~eV is within the band bending of up to about 0.9~eV. This requires the depletion width to approach the escape depth (about 2.5~nm for Si~\cite{S-Miyazaki1999}), which holds for impurity concentrations near $10^{20}$~cm$^{-3}$ and is marginal at $10^{18}$--$10^{19}$~cm$^{-3}$. (ii) Surface and gap states near the Fermi level formed by the HF treatment and the exposure to air; emission from filled gap states below the valence-band threshold has been observed on hydrogen-terminated Si and attributed to slight oxidation during air exposure and rinsing~\cite{S-Miyazaki1999}. This origin is also consistent with the growing deviation from the single power-law model during the series (the decrease of DW in Fig.~3). The exponent $n_1 = 1.0$ was fixed in advance and does not discriminate between these origins: in Kane's theory, states distributed in energy at the Fermi level give $n = 2$ and states localized in energy give $n = 1$ (Table~\ref{S-tab:kane}), whereas with an energy-independent excitation cross section, as assumed by Miyazaki et al.~\cite{S-Miyazaki1999}, a uniform distribution of gap states gives a linear rise. The steep onset of the $n_2 = 2.5$ component from 5.28~eV is consistent with the indirect transition from the valence band near the surface.

In the constrained model, the $n_2 = 2.5$ component is required to set in at or below the threshold of the $n_1 = 1.0$ component. This is the order observed by Gobeli and Allen on a cleaved Si surface, where the direct-transition threshold (5.45~eV) lies above the indirect-transition threshold (5.15~eV), and the order expected if the $n_1 = 1.0$ component were the direct transition from the same valence band. Under this constraint, the optimization places the $n_2 = 2.5$ component at 4.10~eV to follow the early rise and the $n_1 = 1.0$ component at 5.85~eV, where it contributes only at the highest energies; the model reduces effectively to a single component. The AIC is 433.4, an improvement of less than 1, and the undulation of the residuals remains (DW $= 1.08$). Because the constrained and free models have the same number of parameters, the improvement of the free model is a consequence of the order of the thresholds, not of the additional parameters: the rise from 4.70~eV is described by a component with a slow onset below the threshold of the indirect transition. Within the fixed exponents, the direct transition from the valence band near the surface is therefore not supported as the origin of the low-threshold component, and origins (i) and (ii) are not distinguished. The order of the thresholds was not input to the free model but was obtained by the unconstrained optimization.

\section{Criteria and reference values of DW}
\label{S-sec:criteria}

\begin{figure}[p]
\centering
\includegraphics[width=\linewidth]{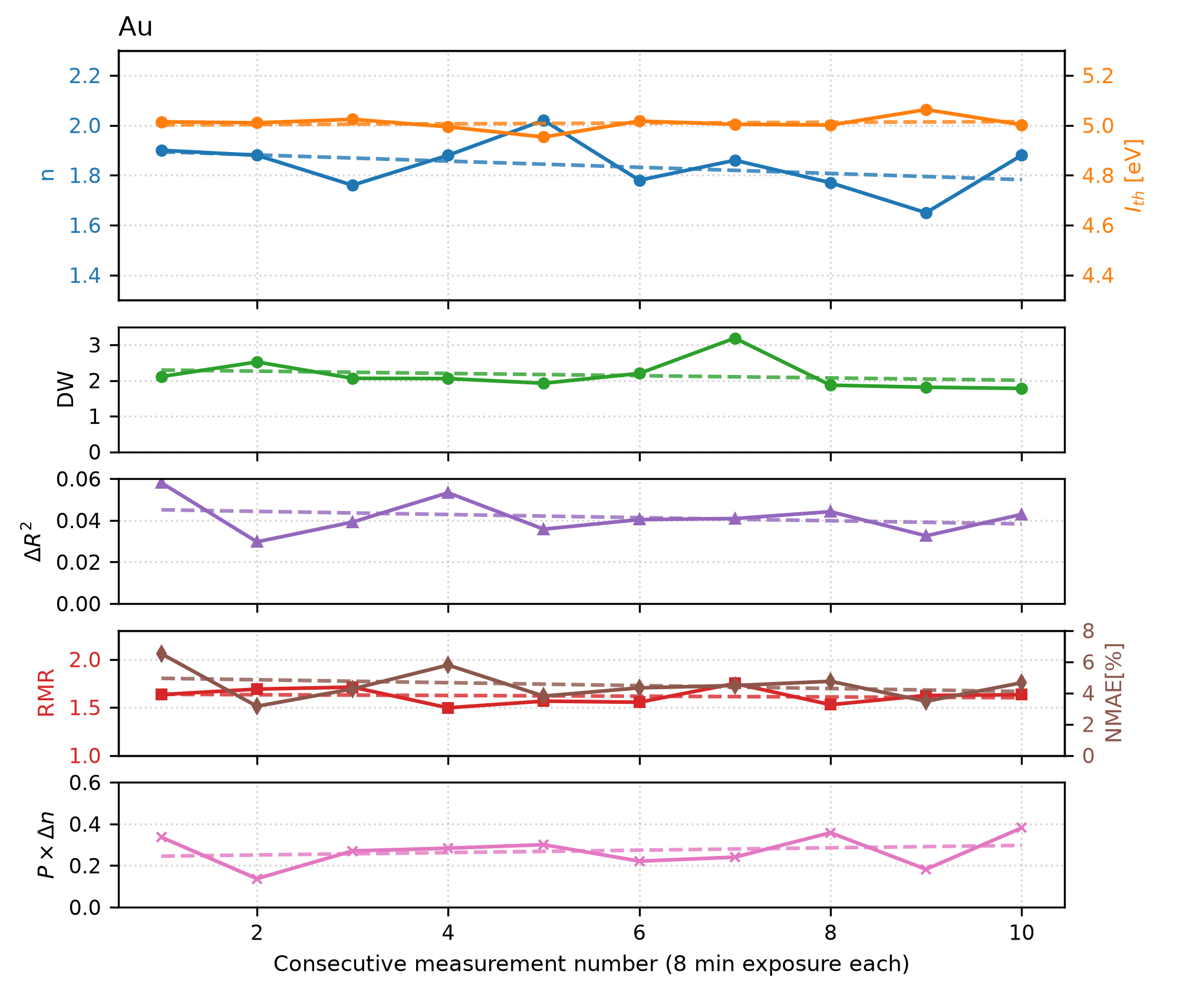}
\caption{Ten continuous measurements of the Au reference. From the top: $n$ (blue, left axis) and $I_{th}$ (orange, right axis); DW; $\Delta R^2$; RMR (red squares, left axis) and NMAE (brown diamonds, right axis); $P \times \Delta n$. Dashed lines are linear trends.}
\label{S-fig:autrend}
\end{figure}

\begin{figure}[p]
\centering
\includegraphics[width=\linewidth,height=0.9\textheight,keepaspectratio]{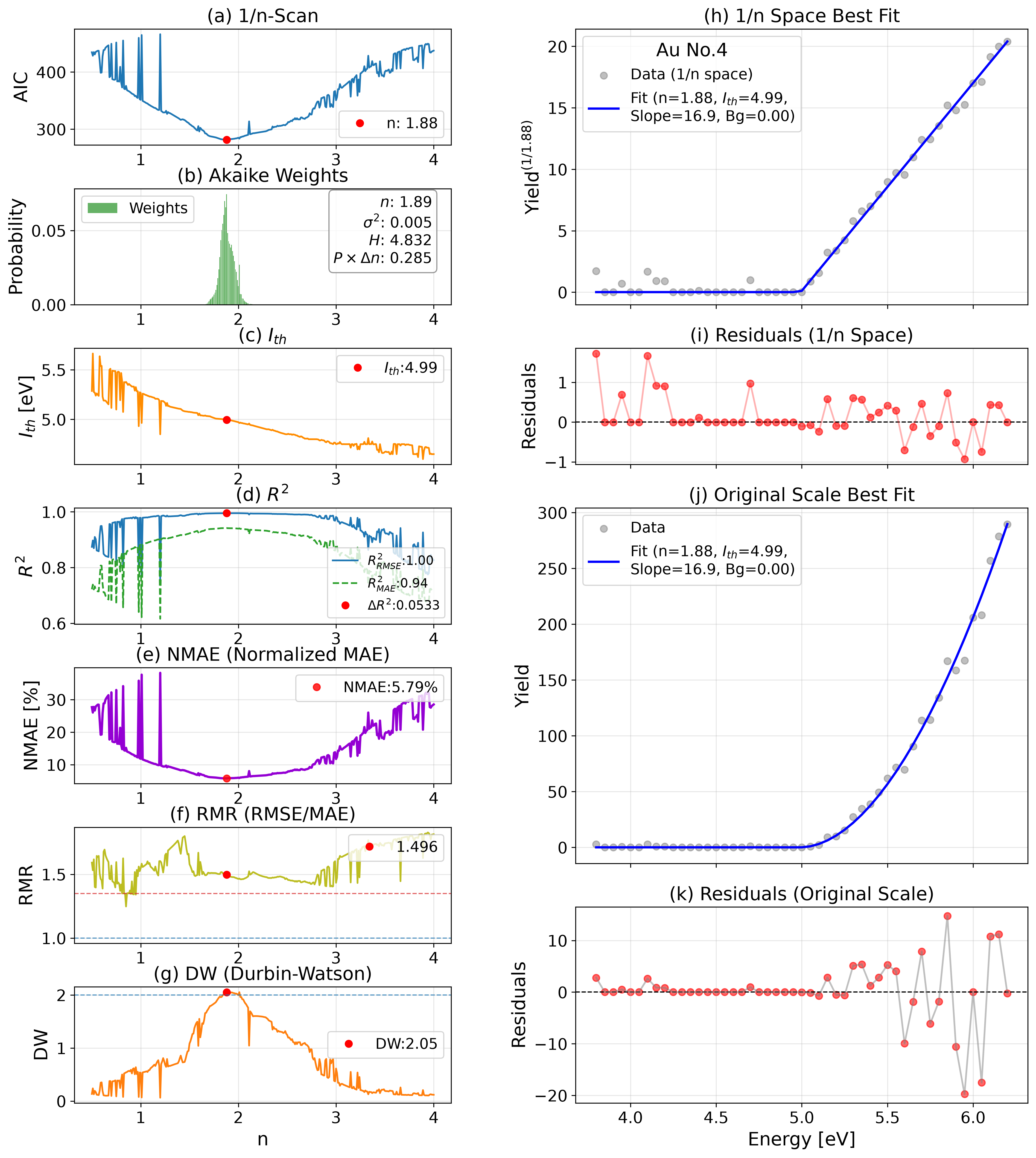}
\caption{Diagnostic output for the fourth measurement of the Au reference (panels as in Fig.~2).}
\label{S-fig:au4}
\end{figure}

\begin{figure}[p]
\centering
\includegraphics[width=\linewidth]{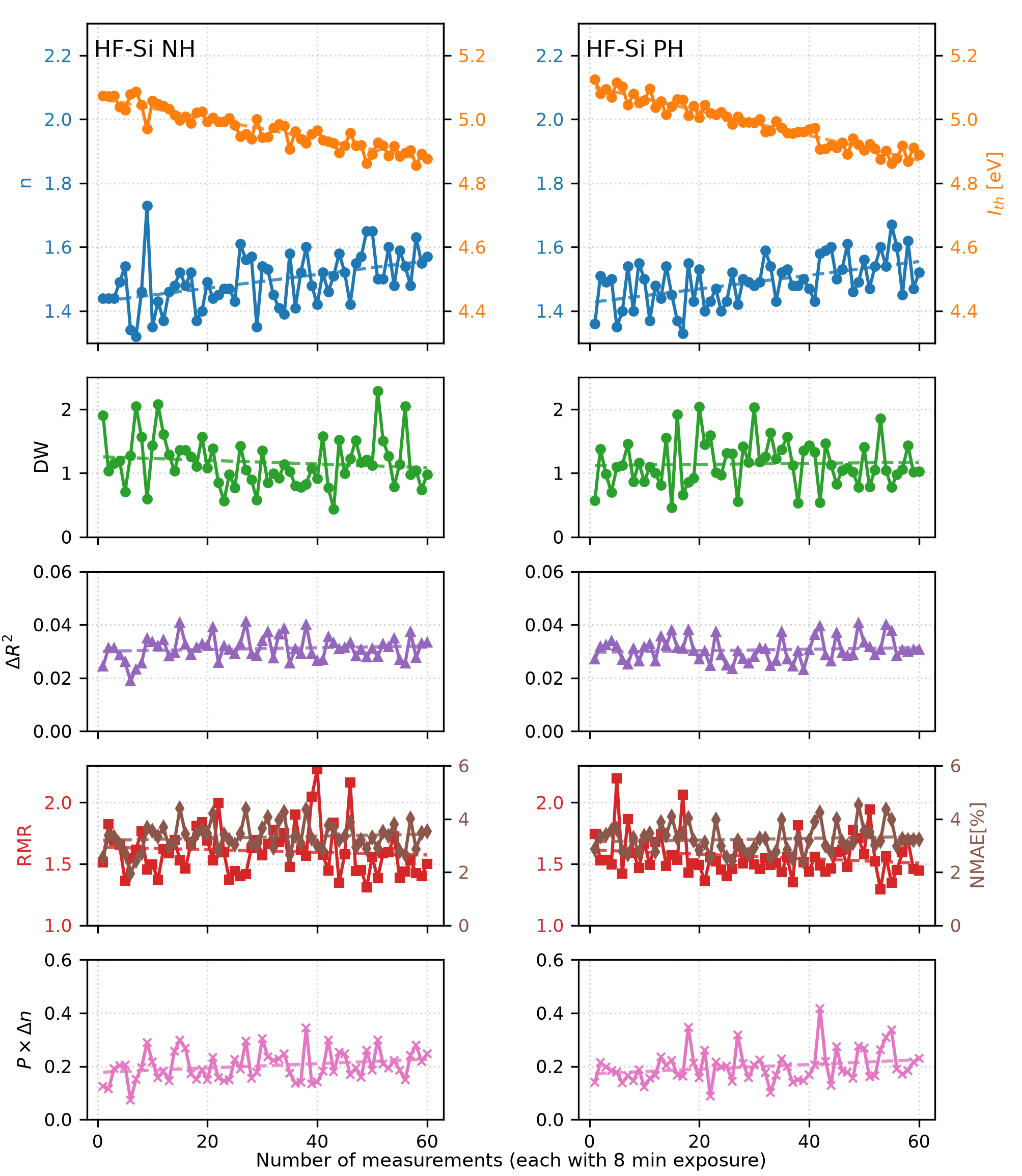}
\caption{Diagnosis of 60 continuous measurements of the HF-treated NH (left) and PH (right) samples. Panels as in Fig.~3: from the top, $n$ (blue, left axis) and $I_{th}$ (orange, right axis); DW; $\Delta R^2$; RMR (red squares, left axis) and NMAE (brown diamonds, right axis); $P \times \Delta n$. $n$ increases gradually from about 1.5 to about 1.6. DW stays at about 1.1--1.2 on average, below $d_L$ from the start of the series, without the progressive decrease of PL.}
\label{S-fig:nhph}
\end{figure}

\begin{figure}[p]
\centering
\includegraphics[width=\linewidth,height=0.6\textheight,keepaspectratio]{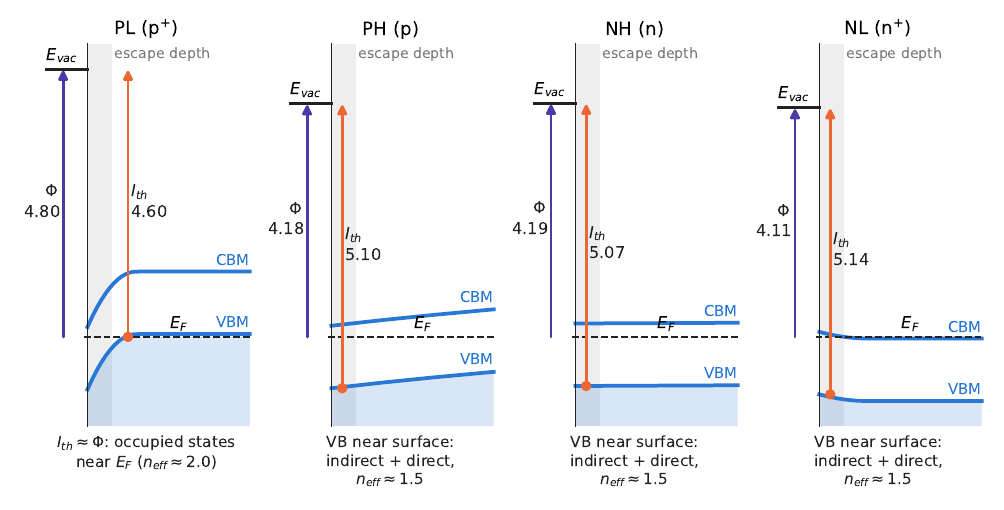}
\caption{Band diagrams of the HF-treated samples at the start of the continuous PYS measurements. $\Phi$ is from CPD series n5 and $I_{th}$ from PYS measurements 1--3 (Table~\ref{S-tab:phiith}), in eV; the depth axis is schematic, and the shaded band marks the escape depth. For PH, NH, and NL the surface VBM is placed at $I_{th} - \Phi$ below $E_F$ (an upper estimate of $E_F - E_{VBM}$); for PL it is placed at the common pinning position obtained by XPS. PH, NH, NL: emission from the valence band near the surface, with overlapping indirect and direct transitions ($n \approx 1.5$). PL: emission from occupied states near $E_F$ ($I_{th} \approx \Phi$); the arrow shows origin (i) of Section~\ref{S-sec:twocomp}, the valence band on the bulk side raised by the band bending; origin (ii), surface and gap states near $E_F$, gives the same threshold.}
\label{S-fig:band}
\end{figure}

\begin{figure}[p]
\centering
\includegraphics[width=\linewidth,height=0.6\textheight,keepaspectratio]{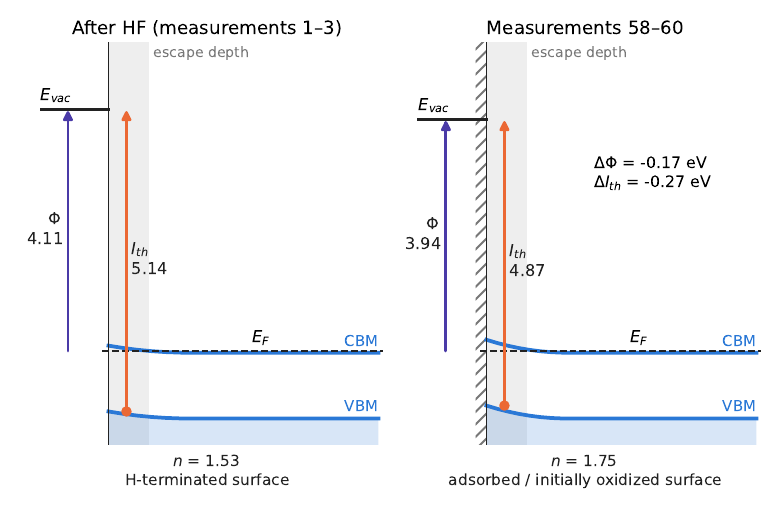}
\caption{Change of NL during the continuous PYS measurements. Left: measurements 1--3; right: measurements 58--60 (CPD after 61 measurements). $\Phi$ and $I_{th}$ decrease by similar amounts, and $I_{th} - \Phi$ is unchanged within 0.1~eV. The increase of $n$ is attributed to the increased scattering of the photoelectrons by the adsorbed and initially oxidized layer (hatched), which reduces the direct-transition component.}
\label{S-fig:nl}
\end{figure}

\begin{figure}[p]
\centering
\includegraphics[width=\linewidth]{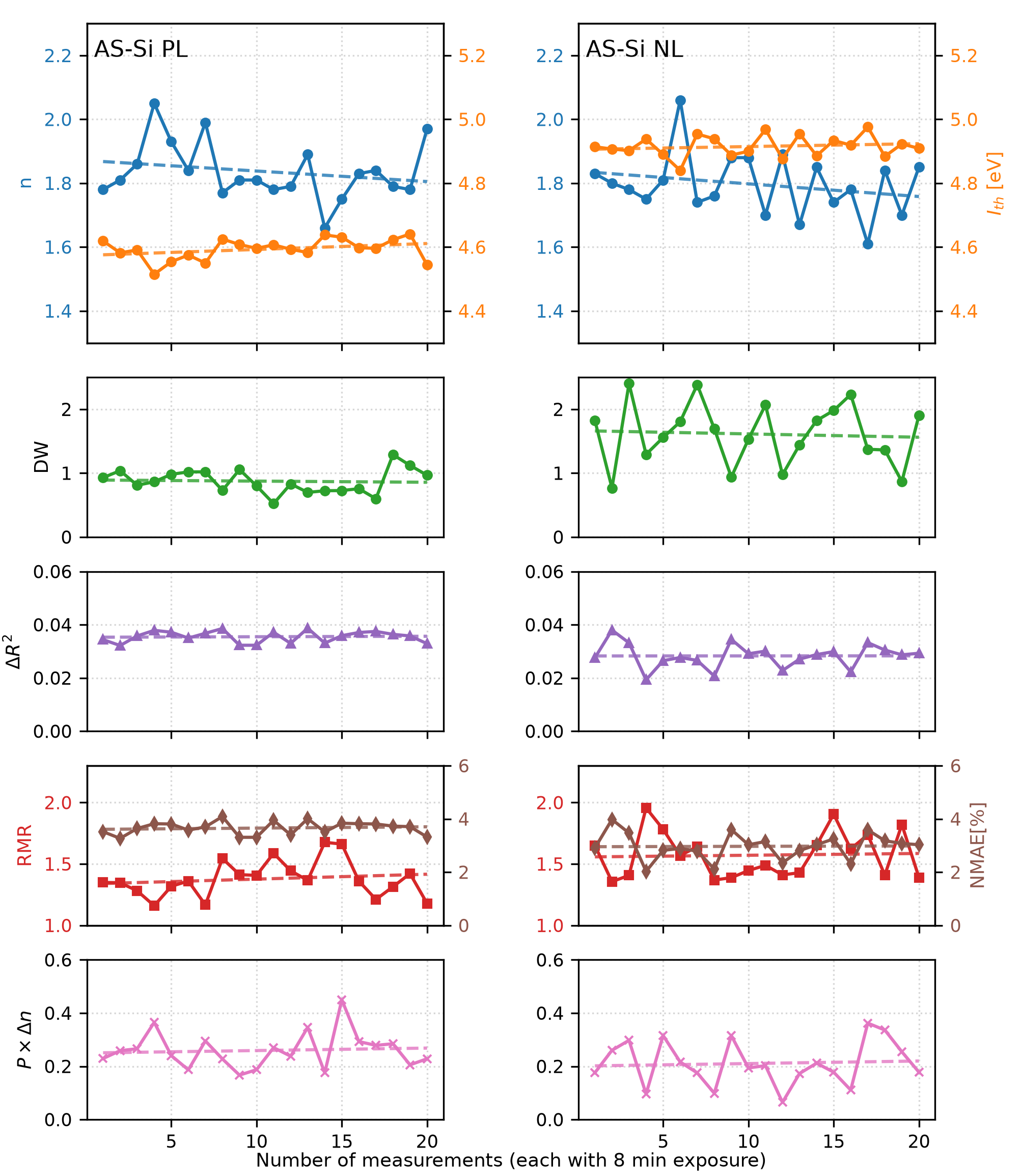}
\caption{$n$ and $I_{th}$ of 20 continuous measurements of the AS heavily doped samples (left: PL; right: NL), measured after the CPD measurement. $n$ is distributed mainly between 1.7 and 2.0 (mean 1.84 for PL and 1.80 for NL).}
\label{S-fig:aspl}
\end{figure}

\begin{figure}[p]
\centering
\includegraphics[width=\linewidth]{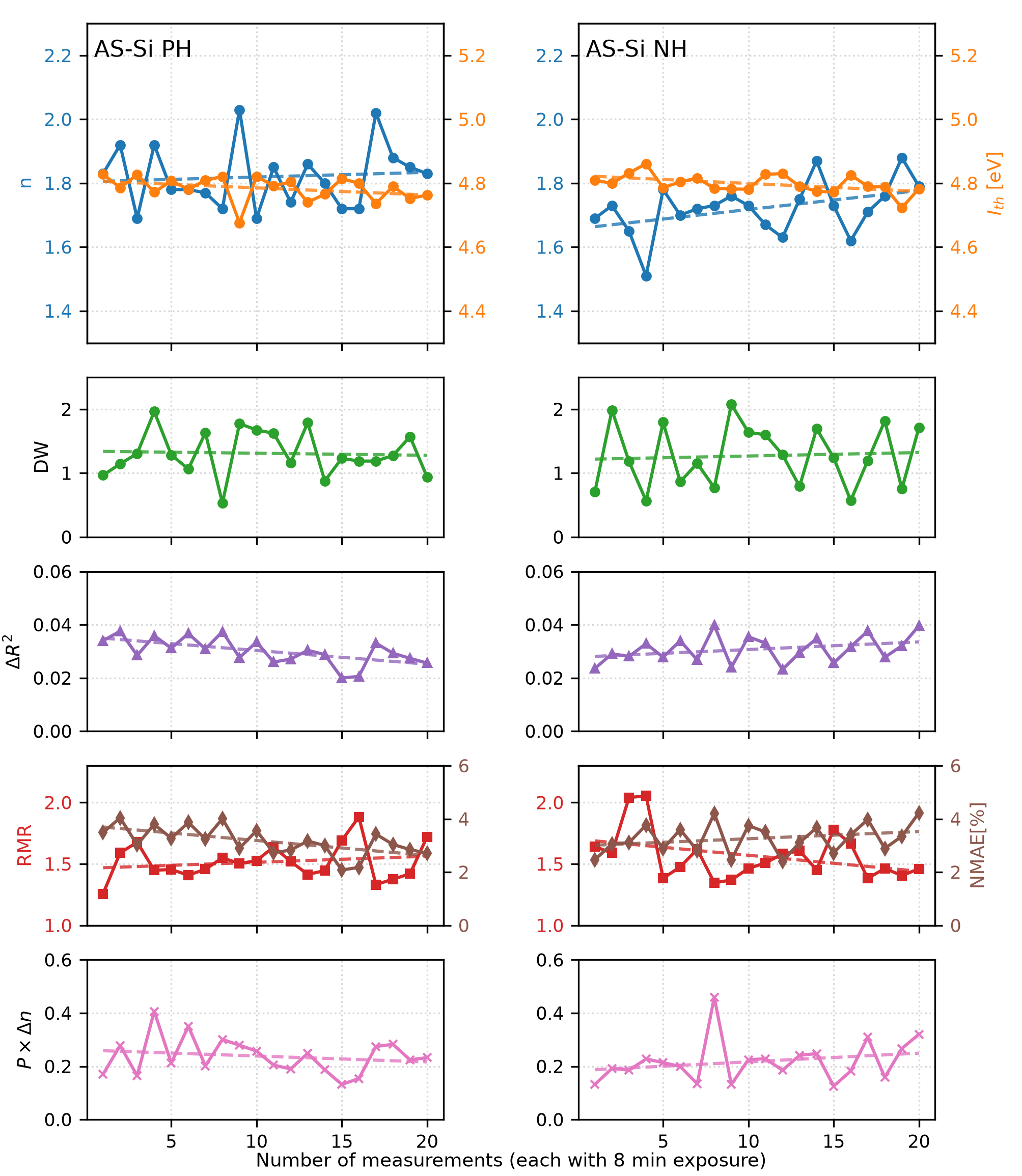}
\caption{$n$ and $I_{th}$ of 20 continuous measurements of the AS lightly doped samples (left: PH; right: NH), measured after the CPD measurement. $n$ is distributed mainly between 1.6 and 2.0 (mean 1.82 for PH and 1.72 for NH).}
\label{S-fig:asph}
\end{figure}

\begin{figure}[p]
\centering
\includegraphics[width=\linewidth,height=0.85\textheight,keepaspectratio]{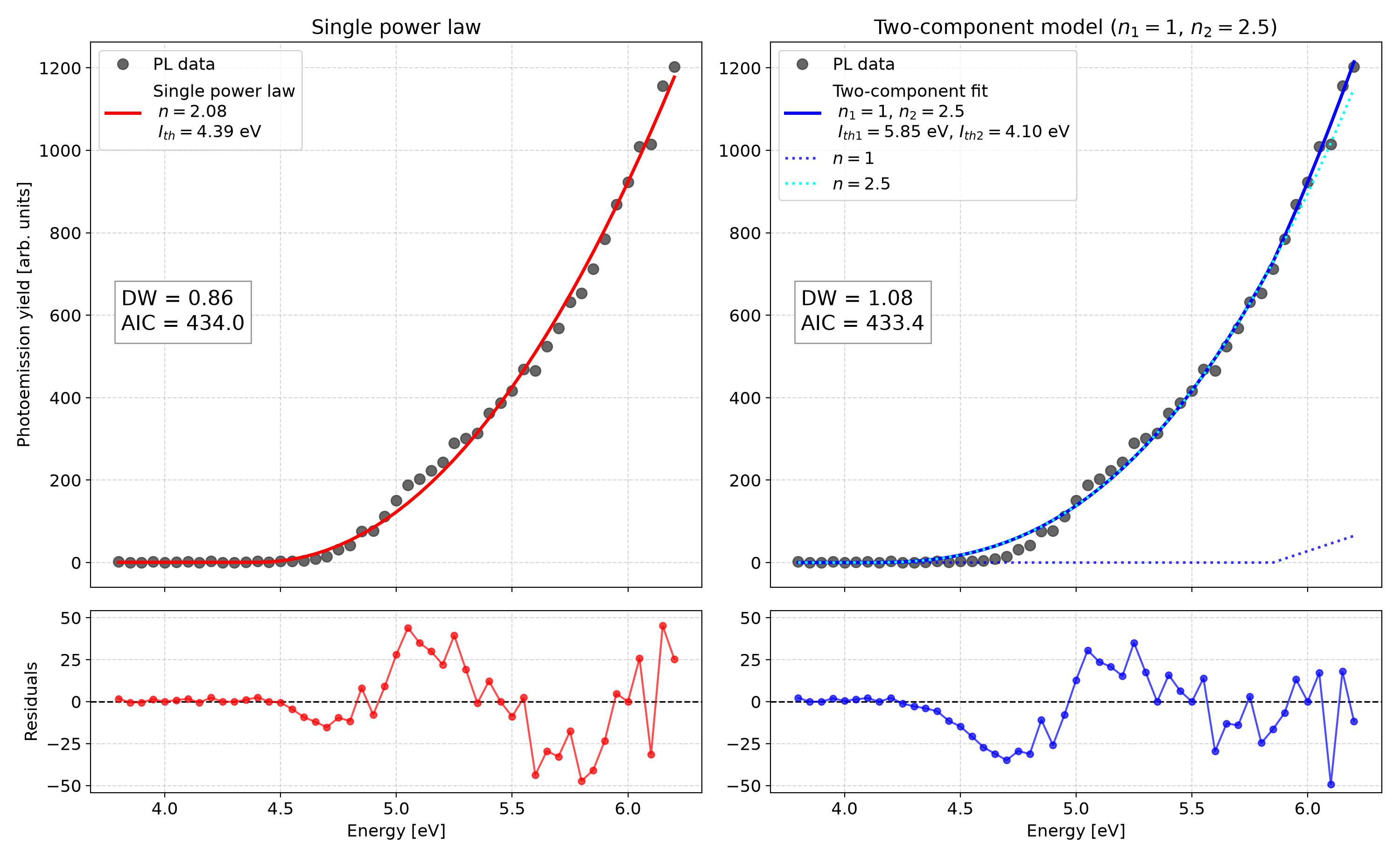}
\caption{Two-component fit of the 30th spectrum of PL with the constraint that the $n_2 = 2.5$ component sets in at or below the threshold of the $n_1 = 1.0$ component (the order of the indirect and direct thresholds of Gobeli and Allen), and its residuals. The optimization places the $n_2 = 2.5$ component at 4.10~eV and the $n_1 = 1.0$ component at 5.85~eV, so that the model is effectively a single component; an undulation remains in the residuals (AIC $= 433.4$, DW $= 1.08$). Left: single power law at $n = 2.08$ (AIC $= 434.0$, DW $= 0.86$).}
\label{S-fig:constrained}
\end{figure}

\textbf{Criteria.} The criteria were set from the 10 continuous measurements of Au (Fig.~\ref{S-fig:autrend}) and are listed in Table~2 of the main text together with the mean values of NL and PL evaluated with them. The residuals of Au show no serial correlation (mean DW 2.15). The criteria were set empirically 5--20\% above the means of Au, and the means of NL lie below them. The criterion RMR $\le 1.7$ was set from the mean of Au (1.62), and the upper warning level RMR $> 2.0$ from the lower end of the range where outliers become apparent. The criterion $P \times \Delta n \le 0.3$ was set from the mean of Au (0.27), and the upper warning level 0.5 from the spacing of Kane's exponents (Section~2.2 of the main text).

\textbf{Reference values of DW.} The lower and upper bounds of the 5\% one-sided test for positive autocorrelation are $d_L \approx 1.46$ and $d_U \approx 1.63$ for $N \approx 50$ and two regressors besides the constant~\cite{S-Durbin1951,S-Savin1977}. As the number of regressors, the two parameters that determine the shape of the residuals above the threshold ($I_{th}$ and the slope) were counted. Counting the background $B'$ as well gives $d_L \approx 1.42$ and $d_U \approx 1.67$. These values were derived for linear least-squares regression with Gaussian errors and do not apply strictly to the present nonlinear fit that minimizes the MAE. DW is therefore used as a diagnostic indicator of serial correlation, not as a formal test, with $d_L$ as the criterion. Only the lower side is used for the check: the overlap of components produces a smooth deviation and therefore positive autocorrelation ($\mathrm{DW} < 2$), whereas a value above $4 - d_L$, as for one Au measurement (DW $= 3.19$, Section~S3), indicates alternating signs of adjacent residuals and is not a sign of model mismatch. DW is preferred to the Breusch--Godfrey test because the deviation of interest is first-order and DW is computed directly from the residual series without an auxiliary regression.

\textbf{DW and RMR under shot noise.} Spectra with the measured $N = 49$ (3.8--6.2~eV), $n = 2.0$, $I_{th} = 5.0$~eV, and a maximum yield of about 700 were generated with noise independent between points and fitted with the $1/n$-power fit at the true $n$ (500 spectra each, NMAE about 3\%). Two noise models were used: uniform Gaussian noise and shot noise with a standard deviation proportional to $\sqrt{Y}$. The results are listed in Table~\ref{S-tab:sim}.

\begin{table}[H]
\centering
\caption{DW and RMR in the simulation. $P(\mathrm{DW} < d_L)$ is the fraction of single spectra with DW below $d_L = 1.46$; the last column is the probability that the median of ten spectra falls below $d_L$ (binomial distribution for independent spectra).}
\label{S-tab:sim}
\small
\begin{tabular}{llcccc}
\toprule
Noise & Residuals & Mean RMR & Mean DW & $P(\mathrm{DW} < d_L)$ & Median of ten \\
\midrule
Uniform Gaussian & $e_i$ & 1.03 & 1.98 & 0.036 & $4\times10^{-7}$ \\
Shot & $e_i$ & 1.55 & 1.97 & 0.138 & $9\times10^{-4}$ \\
Shot & $e_i/\sigma_i$ (true $\sigma_i$) & -- & 2.02 & 0.012 & $6\times10^{-10}$ \\
\bottomrule
\end{tabular}
\end{table}

Under shot noise, the mean of DW stays close to 2 as long as the residuals are independent between points. Because the large residuals concentrate in a few points on the high-yield side, however, the scatter of DW for single spectra increases, and the fraction below $d_L$ rises to 14\%. Standardizing with the true $\sigma_i$ reduces this fraction to 1\%, but in measurements $\sigma_i$ would have to be estimated point by point. In this work, the residuals are not standardized, and continuous measurements are evaluated with the moving median of ten spectra, for which the probability of a false flag is below $10^{-3}$. The mean RMR under shot noise (1.55) is comparable to the RMR measured for Au and NL (1.4--1.6); the excess of the measured RMR over the Laplace value (1.13) is therefore attributable to shot noise.

For NL, individual DW values fell below $d_L$ in 33\% of the measurements (20 of 60), more than the fraction under shot noise (14\%). This is consistent with a weak overlap of the two transitions that does not lower the moving median below $d_L$ (Section~\ref{S-sec:nl}).

\section{Exponents of Kane's theory}

Table~\ref{S-tab:kane} lists the exponents of Kane's Table~I~\cite{S-Kane1962}. The same exponent arises from more than one mechanism: $n = 2$ is obtained for a direct transition in the volume followed by elastic scattering, for an indirect transition from surface band states below the Fermi level, and for imperfection states distributed in energy at the Fermi level; $n = 3/2$ is obtained for a direct transition from a surface band whose threshold coincides with the Fermi level and for specular transitions from volume states at a perfect surface. The exponent alone therefore does not identify the emission mechanism, and the threshold $I_{th}$ is used together with $n$. For the indirect transition, Kane's theory gives $n = 5/2$. Gobeli and Allen stated that the indirect component of clean Si also agrees with an empirical cube law within their experimental accuracy, and determined the indirect threshold (5.15~eV) from an analysis with $n = 5/2$~\cite{S-Gobeli1962}. In the PYS analysis of Si, $n = 3$ is also used~\cite{S-Miyazaki1999}. The two-component fit of this work uses $n = 5/2$, following Gobeli and Allen. In measurements in air, scattering by surface oxides and adsorbates raises the exponent, and when several components overlap, the measured exponent is an effective value.

\begin{table}[H]
\centering
\caption{Exponent $n$ of the photoemission yield $Y \propto (E - E_T)^n$ for the production and scattering mechanisms of Kane's Table~I~\cite{S-Kane1962}. $E_T$ is the threshold and $\mathcal{E}_F$ the Fermi level measured from the vacuum. For a nondegenerate valence maximum at $k = 0$, the Huntington--Apker node adds one power to the surface processes; it is absent for degenerate valence bands such as that of Si.}
\label{S-tab:kane}
\small
\begin{tabular}{p{6.4cm}cp{6.4cm}}
\toprule
Process & $n$ & Condition / origin \\
\midrule
Volume: direct optical excitation, unscattered & 1 & Momentum conserved; no scattering in the volume or at the surface \\
Surface band states: direct excitation, $E_T > |\mathcal{E}_F|$ & 1 & Threshold below the Fermi level \\
Surface imperfection states localized in energy below $\mathcal{E}_F$ & 1 & Discrete defect level \\
Surface band states: direct excitation, $E_T = |\mathcal{E}_F|$ & 3/2 & Metallic surface band crossing the Fermi level \\
Volume states, perfect surface as momentum absorber (specular) & 3/2 & Hole in a volume state; valence maximum along the surface normal (as for Si) \\
Volume: direct optical excitation, elastically scattered & 2 & Scattering in the volume or at the surface before escape \\
Surface band states: indirect excitation, $E_T > |\mathcal{E}_F|$ & 2 & Threshold below the Fermi level \\
Surface imperfection states distributed in energy, $E_T = |\mathcal{E}_F|$ & 2 & Constant density of states at the Fermi level \\
Volume: indirect optical excitation, unscattered or elastically scattered & 5/2 & Phonon- or imperfection-assisted; hole at the valence maximum \\
Volume states, rough surface as momentum absorber & 5/2 & Diffuse surface scattering supplies the momentum \\
Surface band states: indirect excitation, $E_T = |\mathcal{E}_F|$ & 5/2 & Metallic surface band \\
\bottomrule
\end{tabular}
\end{table}

\clearpage

\clearpage

\clearpage

\clearpage

\clearpage

\clearpage

\clearpage

\clearpage

\clearpage


\begin{thebibliography}{99}
\bibitem{Kane1962} Kane EO. Theory of photoelectric emission from semiconductors. Phys Rev. 1962;127(1):131--141. doi:10.1103/PhysRev.127.131
\bibitem{Gobeli1962} Gobeli GW, Allen FG. Direct and indirect excitation processes in photoelectric emission from silicon. Phys Rev. 1962;127(1):141--149. doi:10.1103/PhysRev.127.141
\bibitem{Miyazaki1999} Miyazaki S, Maruyama T, Kohno A, Hirose M. Photoelectron yield spectroscopy of electronic states at ultrathin interfaces. Microelectron Eng. 1999;48:63. doi:10.1016/S0167-9317(99)00339-1
\bibitem{Yagyu2025} Yagyu S, Yoshitake M, Nagata T, Yasuda T, Wakahara T, Liu Y, Nakajima Y. Automated analysis of photoelectron yield spectroscopy spectra interpreted via power laws. Sci Technol Adv Mater Meth. 2025;5(1):2465257. doi:10.1080/27660400.2025.2465257
\bibitem{Yagyu2026} Yagyu S, Nagata T, Yasuda T, Nakajima Y. Robust and automated analysis framework for photoelectron yield spectroscopy: overcoming statistical bias via MAE-based AIC and log-log linearization. Sci Technol Adv Mater Meth. In press. doi:10.1080/27660400.2026.2744551
\bibitem{Miyawaki2026} Miyawaki N, Hitosugi T. Materials science in the era of AI: recent trends in autonomous experimental systems and the changing role of scientists. Appl Phys Express. 2026;19:070102. doi:10.35848/1882-0786/ae7176
\bibitem{Durbin1951} Durbin J, Watson GS. Testing for serial correlation in least squares regression. II. Biometrika. 1951;38(1--2):159--178. doi:10.1093/biomet/38.1-2.159
\bibitem{Savin1977} Savin NE, White KJ. The Durbin-Watson test for serial correlation with extreme sample sizes or many regressors. Econometrica. 1977;45:1989. doi:10.2307/1914122
\bibitem{Fowler1931} Fowler RH. The analysis of photoelectric sensitivity curves for clean metals at various temperatures. Phys Rev. 1931;38(1):45--56. doi:10.1103/PhysRev.38.45
\bibitem{Burnham2004} Burnham KP, Anderson DR. Multimodel inference: understanding AIC and BIC in model selection. Sociol Methods Res. 2004;33:261--304. doi:10.1177/0049124104268644
\bibitem{Sze2006} Sze SM, Ng KK. Physics of semiconductor devices. 3rd ed. Hoboken (NJ): Wiley; 2006. doi:10.1002/0470068329
\bibitem{Grzibovskis2018} Grzibovskis R, Vembris A. Energy level determination in bulk heterojunction systems using photoemission yield spectroscopy: case of P3HT:PCBM. J Mater Sci. 2018;53:7506. doi:10.1007/s10853-018-2050-9
\bibitem{Grzibovskis2016} Grzibovskis R, Vembris A, Pudzs K. Relation between molecule ionization energy, film thickness and morphology of two indandione derivatives thin films. J Phys Chem Solids. 2016;95:12. doi:10.1016/j.jpcs.2016.03.010
\bibitem{Kochi1970} Kochi M, Harada Y, Hirooka T, Inokuchi H. Photoemission from organic crystal in vacuum ultraviolet region. IV. Bull Chem Soc Jpn. 1970;43(9):2690--2702. doi:10.1246/bcsj.43.2690
\bibitem{Yagyu2022} Yagyu S. Program for extracting metadata from measurement files of the photoelectron yield spectrometer in air (AC series by RIKEN KEIKI Co., Ltd.). J Surf Anal. 2022;29:97. doi:10.1384/jsa.29.97
\end{thebibliography}

\begin{thebibliography}{9}
\bibitem{S-Irvin1968} Sze SM, Irvin JC. Resistivity, mobility and impurity levels in GaAs, Ge, and Si at 300~K. Solid-State Electron. 1968;11:599--602. doi:10.1016/0038-1101(68)90012-9
\bibitem{S-Miyazaki1999} Miyazaki S, Maruyama T, Kohno A, Hirose M. Photoelectron yield spectroscopy of electronic states at ultrathin interfaces. Microelectron Eng. 1999;48:63. doi:10.1016/S0167-9317(99)00339-1
\bibitem{S-Gobeli1962} Gobeli GW, Allen FG. Direct and indirect excitation processes in photoelectric emission from silicon. Phys Rev. 1962;127(1):141--149. doi:10.1103/PhysRev.127.141
\bibitem{S-Kane1962} Kane EO. Theory of photoelectric emission from semiconductors. Phys Rev. 1962;127(1):131--141. doi:10.1103/PhysRev.127.131
\bibitem{S-Sze2006} Sze SM, Ng KK. Physics of semiconductor devices. 3rd ed. Hoboken (NJ): Wiley; 2006. doi:10.1002/0470068329
\bibitem{S-Durbin1951} Durbin J, Watson GS. Testing for serial correlation in least squares regression. II. Biometrika. 1951;38(1--2):159--178. doi:10.1093/biomet/38.1-2.159
\bibitem{S-Savin1977} Savin NE, White KJ. The Durbin-Watson test for serial correlation with extreme sample sizes or many regressors. Econometrica. 1977;45:1989. doi:10.2307/1914122
\end{thebibliography}
\end{document}